\documentclass[a4paper, 12pt, reqno]{amsart}
\usepackage{amsmath}
\usepackage{amsfonts}
\usepackage{graphics}
\usepackage{epsfig}
\usepackage{amssymb}
\usepackage{amscd}
\usepackage[all]{xy}
\usepackage{latexsym}
\usepackage{graphicx}
\usepackage{multirow}
\usepackage{geometry}
\usepackage{color}
\usepackage{hyperref}
\usepackage{url}
\usepackage{color}

 \newtheorem{thm}{Theorem}[section]

 \newtheorem{prop}[thm]{Proposition}
 \theoremstyle{definition}
 \newtheorem{defn}[thm]{Definition}
 \theoremstyle{remark}
 \newtheorem{rem}[thm]{Remark}
 
 \numberwithin{equation}{section}

\def\to{\rightarrow}
\def\td{\tilde}
\def\hs{\hspace}
\def\vs{\vspace}

\def\Tr{{\rm Tr}}

\def\RR{\mathcal R}
\def\beq{\begin{equation}}
\def\eeq{\end{equation}}
\def\bea{\begin{eqnarray}}
\def\eea{\end{eqnarray}}

\newcommand{\Res}{\mathop{\rm Res}}

\begin{document}

%
%
%
%
%
%
%
%
%

\title[Painlev\'e $2$, topological 
recursion and determinantal formulas]
{Painlev\'e $2$ equation with arbitrary monodromy parameter, 
topological recursion and determinantal formulas}

\author[Kohei Iwaki]{Kohei Iwaki}

\address{%
Graduate School of Mathematics \\
Nagoya University \\
Japan}

\email{iwaki@math.nagoya-u.ac.jp}


\author{Olivier Marchal}
\address{Universit\'e de Lyon \\ 
CNRS UMR 5208 \\ 
Universit\'e Jean Monnet \\ 
Institut Camille Jordan \\ France}
\email{olivier.marchal@univ-st-etienne.fr}


\date{\today}

\begin{abstract}
The goal of this article is to prove that the determinantal formulas of the Painlev\'e $2$ system identify with the correlation functions computed from the topological recursion on their spectral curve for an arbitrary non-zero monodromy parameter. The result is established for a WKB expansion of two different Lax pairs associated to the Painlev\'e $2$ system, namely the Jimbo-Miwa Lax pair and the Harnad-Tracy-Widom Lax pair, where a small expansion parameter $\hbar$ is introduced by a proper rescaling. The proof is based on showing that these systems satisfy the topological type property introduced in \cite{BBEnew, Deter}. In the process, we explain why the insertion operator method traditionally used to prove the topological type property is currently incomplete and we propose new algebraic methods to bypass the issue. Our work generalizes similar results obtained from random matrix theory in the special case of vanishing monodromies \cite{P2WithoutCoeffFull, P2WithoutCoeff}. Explicit computations up to $g=3$ are provided along the paper as an illustration of the results. Eventually, taking the time parameter $t$ to infinity we observe that the symplectic invariants $F^{(g)}$ of the Jimbo-Miwa and Harnad-Tracy-Widom spectral curves converge to the Euler characteristic of moduli space of genus $g$ Riemann surfaces. 
\end{abstract}

\maketitle
\tableofcontents

\section{Introduction}

In the past decade, the connection between random matrix theory, topological recursion and integrable systems has been developed intensively. Indeed, it was first proved that the partition function describing Hermitian random matrix models (first one matrix models and later two matrix models) are isomonodromic tau-functions \cite{Iso1,Iso2}, a central element of integrable systems. Additionally, the local statistics of eigenvalues in Hermitian matrix models have been proved to be universal and related to Fredholm determinants whose kernels are determined by the nature of the point in the global distribution (edge, bulk point, critical points, etc.) \cite{MehtaBook, TW}. Lately, these Fredholm determinants were expressed with some Painlev\'e transcendents \cite{TW}. Recently, Eynard and Orantin provided a recursive algorithm, known as ``the topological recursion'' \cite{EO}, to compute the (possibly formal) $1/N$ expansion of the correlation functions and partition function of any Hermitian matrix model. This recursion was generalized almost immediately to any ``spectral curve'' that may or may not come from a matrix model. This topological recursion has been proved very useful in enumerative geometry where many combinatorial results were recently obtained or rediscovered with this formalism \cite{Dessin, Panda, ListSpecCurve, IntersTheory}. In particular the main interest of the topological recursion is the definition of a series of numbers $F^{(g)}$ known as ``symplectic invariants'' that are invariant under a certain class of symplectic transformations of the initial spectral curve and that reconstruct the logarithm of the partition function when the spectral curve arises from a matrix model. In a more recent article, Berg\`ere and Eynard \cite{Deter} were able to associate a natural spectral curve to any $2\times 2$ Lax pair and provided some determinantal formulas attached to the Lax pair. These determinantal formulas match the correlation functions and symplectic invariants obtained from the computation of the topological recursion on the spectral curve when some additional conditions, known generically as the ``topological type'' (TT) property, are satisfied. More recently, these notions were extended successfully to $n\times n$ Lax pairs by Berg\`ere, Borot and Eynard \cite{BBEnew}. These results are important since they can be used to prove that the determinantal formulas and the tau-function of the Lax pair can be computed perturbatively to any order with the topological recursion associated to the spectral curve, which in general is relatively easy. So far, the TT property has been discussed in three different cases:
\begin{itemize}\item First in \cite{BE2}, in relation with the local statistics of eigenvalues near the edge of the distribution for a Hermitian matrix model, the authors proved the TT property for the Painlev\'e $2$ system (with the Jimbo-Miwa Lax pair) with vanishing monodromy. The approach was generalized in the case of a critical edge with the $(2m,1)$ hierarchy in \cite{MC}. These results where recently recovered and precised in \cite{P2WithoutCoeffFull, P2WithoutCoeff}.
\item In \cite{P5}, in relation with local statistics of eigenvalues in the bulk of the distribution for a Hermitian matrix model, the authors proved that the result holds for the Painlev\'e $5$ system with vanishing monodromy parameters. 
\item Eventually in \cite{BBEnew}, the authors were able to prove the TT property for the $q$-th reduction of the KP hierarchy, that is to say all $(p,q)$ models. In particular this includes the Painlev\'e $1$ equation (for which there is no monodromy parameter).
\end{itemize}
However it is worth mentioning that all these articles use at some point the method of the insertion operator. In this article we show that the current proof regarding the insertion operator is incomplete and we present another way based on loop equations to bypass this issue (see Section \ref{AppendixC} and Appendix \ref{Incomplete}).

\medskip

In this article, our main goal is to prove that the TT property holds for
Lax pairs of the Painlev\'e $2$ equation:
\begin{equation} \label{eq:painleve2-intro}
\hbar^{2} \ddot{q} = 2 q^3 + t q - \theta + \frac{\hbar}{2} 
\end{equation}
(where $\dot{} = \frac{d}{dt}$ and $\hbar$ is a small parameter) with arbitrary non-zero monodromy parameter $\theta$. More precisely, we will prove the result for two different ($\hbar$-deformed) Lax pairs frequently used to describe the Painlev\'e $2$ system: the Jimbo-Miwa (JM) Lax pair and the Harnad-Tracy-Widom (HTW) Lax pair. (See \cite{HTW, JMII}.) For these two Lax pairs we will review how to produce the spectral curve and the tau-function. Although these Lax pairs describe the same integrable system (Painlev\'e $2$), their spectral curves are totally different. Then, after presenting the topological recursion and the determinantal formulas, we will prove the TT property by proving the three conditions proposed in \cite{Deter}. 

This result proves that the generating functions for both sets of symplectic invariants $F^{(g)}_{\rm JM}(t)$ and $F^{(g)}_{\rm HTW}(t)$ defined from the spectral curves of JM pair and HTW pair give the corresponding tau-functions of Painlev\'e 2 (see Theorem \ref{thm:JMFg-tau} and \ref{FgHTW}). Note that, although $F^{(g)}_{\rm JM}(t)$ and $F^{(g)}_{\rm HTW}(t)$ are different as a computational result, both of them gives tau-functions of Painlev\'e 2 equation. As presented in Theorem \ref{propo3} and \ref{thm:Bessel-Fg} the connection between the two sets of symplectic invariants appears from constant terms (a tau-function is always defined up to a constant) that are fixed by the topological recursion. More specifically one of our main results is that:
\begin{equation} \label{eq:difference-Fg-intro}
F^{(g)}_{\rm JM}(t)=F^{(g)}_{\rm HTW}(t) - \frac{B_{2g}}{2g(2g-2)\theta^{2g-2}} \quad \text{for $g \ge 2$}. 
\end{equation}
Along the article $B_{2g}$ stands for the Bernoulli numbers defined by
\begin{eqnarray*} 
\frac{t}{e^{t}-1}-1+\frac{t}{2}=\sum_{m=1}^{\infty}B_{2m}\frac{t^{2m}}{(2m)!}.  \end{eqnarray*}
We could find the reason why the specific constant terms appear and we connect them to two simple spectral curves: the Hermite-Weber curve (semi-circle curve) for which the symplectic invariants have been known and the Bessel curve for which we could not find any reference in the literature. Actually, the JM curve and HTW curve are related by a  symplectic transformation (see Remark \ref{remark:Yamada}; we thank Professor Yasuhiko Yamada who suggests this fact to the authors). We also show that the above discrepancy between $F^{(g)}$'s is consistent with the integration constant computed in \cite{EO-xy-symmetry} (see Appendix \ref{AppendixE}).


\section{Summary of the main results}
This article aims at a better understanding between the integrable structure of the Painlev\'{e} equations and the topological recursion. Our main results are:
\begin{itemize} 
\item We prove that the determinantal formulas and tau-functions associated to two different Painlev\'e $2$ Lax pairs (Jimbo-Miwa and Harnad-Tracy-Widom) are identical to the correlation functions and symplectic invariants computed by the topological recursion applied to the corresponding spectral curves (Theorem \ref{thm:JMFg-tau}, \ref{FgHTW} and \ref{MainTheo}). Explicit results for the expansion of the tau-function are also presented for both Lax pairs. 
\item 
We also show that some limit $t \to \infty$ of the symplectic invariants of both Lax pairs coincide with the Euler characteristic of the moduli space ${\mathcal M}_g$ of Riemann surfaces of genus $g$ computed in \cite{HZ, Penner}, up to sign (Theorem \ref{propo3} and \ref{thm:Bessel-Fg}). This also prove the constant term in the relation \eqref{eq:difference-Fg-intro} between two symplectic invariants for JM and HTW spectral curves (Appendix \ref{AppendixE}). 
\item New methods of proof of the TT property are introduced in this article (Section \ref{AppendixC}). In particular the presence of a compatible time differential system is shown to be of critical importance to the TT property (Appendix \ref{AppendixA}). Moreover these new methods can be easily applied to more general situations and should provide a way to perform the same analysis for the other Painlev\'{e} equations.
\item We show that the method of the insertion operator used in several papers is incomplete since there is a subtle gap in the proof (Appendix \ref{Incomplete}). This issue was the main reason for the introduction of new methods to prove the TT property.   
\end{itemize}


\section{Jimbo-Miwa Lax pair for Painlev\'e $2$}

In this section 
we present the ($\hbar$-depending) Jimbo-Miwa Lax pair as well as the tau-function and its expansion in $\hbar$. Finally we compute the spectral curve and we illustrate our results with the computation of the first few symplectic invariants $F^{(g)}$ defined by performing the topological recursion, and compare them to the tau-function.

\subsection{\label{GaugeChoice}Introduction of $\hbar$ in the Jimbo-Miwa Lax pair}

The following $2\times 2$ Lax pair is equivalent (up to a certain gauge transformation) to the one given in Appendix C of \cite{JMII}:
\begin{equation} \label{eq:original-JM}
\left\{
\begin{array}{lll}
\displaystyle
\frac{\partial \Psi}{\partial x}(x,t)&=&
\begin{pmatrix} 
\displaystyle
x^2+p+\frac{t}{2}&
\displaystyle x-q \\
\displaystyle -2\left(xp+qp+\theta\right) &
\displaystyle -\left(x^2+p+\frac{t}{2}\right)
\end{pmatrix} \Psi(x,t), \\[+.3em]
\displaystyle \frac{\partial \Psi}{\partial t}(x,t) &=& 
\begin{pmatrix} \displaystyle \frac{x+q}{2} & 
\displaystyle \frac{1}{2}\\
\displaystyle -p& \displaystyle -\frac{x+q}{2}
\end{pmatrix} \Psi(x,t).
\end{array} 
\right.
\end{equation}

The Lax pair we will use in this article is a $\hbar$-deformed version of \eqref{eq:original-JM}. The introduction of $\hbar$ is essential in the relation to topological recursion, and it is done by the following rescaling of all quantities involved in the former Lax pair: 
\begin{eqnarray} 
x \to \hbar^{-\frac{1}{3}}\td{x} \,\,,\,\, 
p\to \hbar^{-\frac{2}{3}}\td{p} \,\,,\,\,  
t \to \hbar^{-\frac{2}{3}}\td{t}\,\,,\,\,  
q\to \hbar^{-\frac{1}{3}}\td{q}\,\,,\,\, 
\theta\to \hbar^{-1}\td{\theta}. 
\end{eqnarray}
The scaling degrees are chosen so that the resulting Lax pair can be treated in terms of WKB method. The above rescaling provides the $\hbar$-deformed Lax pair (we omit $\tilde{~}$ for clarity):
\begin{eqnarray} 
\label{JMLaxPair}\left\{
\begin{array}{lll}
\displaystyle 
\hbar\frac{\partial \Psi}{\partial x}(x,t)&=&
\begin{pmatrix} 
\displaystyle x^2+p+\frac{t}{2}&x-q\\
-2\left(xp+qp+\theta\right)& -
\displaystyle \left(x^2+p+\frac{t}{2}\right)
\end{pmatrix} \Psi(x,t) \\ 
& \hspace{-0.1em} 
\overset{\text{def}}{=} & \mathcal{D}(x,t) \Psi(x,t), \\[+.7em]
\displaystyle \hbar\frac{\partial \Psi}{\partial t}(x,t)&=&
 \begin{pmatrix} 
 \displaystyle \frac{x+q}{2}& \displaystyle  \frac{1}{2}\\
-p& \displaystyle -\frac{x+q}{2}\end{pmatrix} \Psi(x,t) \\
& \hspace{-0.1em} \overset{\text{def}}{=} & 
\mathcal{R}(x,t) \Psi(x,t).
\end{array}
\right.
\end{eqnarray}
We call the Lax pair \eqref{JMLaxPair} the Jimbo-Miwa pair (JM pair, for short).

We remind the reader that $q$ and $p$ are implicitly assumed to depend on the time variable $t$ (and also on $\hbar$) but not on $x$. Moreover $\theta$, called the monodromy parameter, is independent of $x,t$ and $\hbar$. Throughout of the paper, we assume that
\begin{equation} 
\label{eq:assume-theta} \theta \ne 0. 
\end{equation}

The compatibility equations of the differential system (also known as zero-curvature equations) are given by:
\begin{equation} 
\hbar \left(\frac{\partial \mathcal{D}}{\partial t} -\frac{\partial \mathcal{R}}{\partial x}\right)+\left[\mathcal{D},\mathcal{R}\right]=0. 
\end{equation}
From \eqref{JMLaxPair} they are equivalent to:
\begin{equation} 
\label{DeformedEq}\hbar \dot{p}=-2qp-\theta \,\,,\,\, 
\hbar \dot{q}=p+q^2+\frac{t}{2}. 
\end{equation}
Here and in what follows a dot is used to denote the derivative relatively to $t$ when no ambiguity appears. 
Differentiating the last equation and eliminating $p$ with the first equation gives that $q$ satisfies the Painlev\'e $2$ equation (with the small parameter $\hbar$):
\begin{equation}
 \label{Pain}\hbar^2 \ddot{q}=2q^3+tq-\theta+\frac{\hbar}{2}. 
\end{equation} 
This type of Painlev\'e equations, and associated Lax pairs with a small parameter $\hbar$ were studied in \cite{Kawai}. 

In this paper we are interested in the $\hbar$-perturbative expansion of a solution of Painlev\'e $2$ equation:
\begin{equation} \label{eq:formal-solution-q}
q(t)=\sum_{k=0}^\infty q_k(t)\hbar^k=q_0(t)+q_1(t) 
\hbar+q_2(t)\hbar^2+\cdots. 
\end{equation}
The top term $q_{0}(t)$ satisfies 
\begin{equation} 
\label{y0JM}2q_0(t)^3+tq_0(t)-\theta=0 \text{ and } 
\dot{q_0}(t)=-\frac{q_0(t)^2}{4q_0(t)^3+\theta}. 
\end{equation}
As is clear from \eqref{y0JM}, $\dot{q}_{0}(t)$ has singularity where $4q_0(t)^3+\theta = 0$ holds. Such a point on $t$-plane is called a turning point of Painlev\'e 2 in \cite{Kawai}. In what follows, we assume that $t$ lies on a domain on which $4q_0(t)^3+\theta \ne 0$ holds. Note also that $q_{0}(t) \ne 0$ holds for any $t$ under the assumption \eqref{eq:assume-theta}. 

Since $q_0(t)$ is a solution of a cubic equation, there are $3$ possible choices of branches for $q_0(t)$. In particular when $t\to \infty$ there are three possible behaviors depending on the chosen branch:
\begin{equation} 
q_0(t)\underset{t\to \infty_A}{\sim} 
\frac{\theta}{t}\,\,,\,\, ~~~~
q_0(t)\underset{t\to \infty_B}{\sim} \sqrt{\frac{-t}{2}}\,\,,\,\, ~~~~
q_0(t)\underset{t\to \infty_C}{\sim} -\sqrt{\frac{-t}{2}}. 
\end{equation}
It is easy to see that, once we fix the branch of the algebraic function $q_{0}(t)$, the coefficients $\{ q_{i}(t) \}_{i \ge 1}$ appearing in \eqref{eq:formal-solution-q} are determined recursively. Thanks to the relation \eqref{DeformedEq}, we also get a similar $\hbar$-expansion of $p(t)$.  

\subsection{Hamiltonian system and tau-function of the JM Lax pair}

The tau-function are classically defined since the works of Jimbo-Miwa-Ueno \cite{JMI}. For the JM pair, these quantities are easy to derive (the leading order of the matrices $\mathcal{D}(x,t)$ and $\mathcal{R}(x,t)$ when $x\to \infty$ are both diagonal) and can be directly adapted from the known $\hbar=1$ case. The Hamiltonian system attached to the JM pair \eqref{JMLaxPair} is:
\begin{eqnarray} 
\label{eq:Ham2-JM} \left\{
\begin{array}{lll}
\displaystyle
\hbar \dot{q} &=& \displaystyle
\frac{\partial H_{\rm JM}}{\partial p}=
p +q^2+\frac{t}{2},\\[+.7em]
\displaystyle
\hbar \dot{p}&=& \displaystyle
-\frac{\partial H_{\rm JM}}{\partial q}=
-2qp -\theta,\\
\end{array}\right.
\end{eqnarray}
where $H_{\rm JM}$ is the Hamiltonian for Painlev\'e 2:
\begin{equation}
H_{\rm JM} = \frac{1}{2}p^2+\left(q^2+\frac{t}{2}\right)p+\theta q.
\end{equation} 
Let $\sigma(t)$ be the corresponding Hamiltonian function, that is, the function obtained by substituting a solution $(q,p)$ of \eqref{eq:Ham2-JM} into $H_{\rm JM}$. It satisfies:
\begin{equation}  \label{relation}
\dot{\sigma}=\frac{p}{2} \text{ and } 
\hbar\ddot{\sigma}=-qp-\frac{\theta}{2}.
\end{equation}
as well as the $\sigma$-form of the Painlev\'e $2$ equation:
\begin{equation}  \label{eqtau}
\left(\hbar\ddot{\sigma}\right)^2+4\left(\dot{\sigma}\right)^3+2t\left(\dot{\sigma}\right)^2-2\sigma\dot{\sigma}-\frac{\theta^2}{4}=0.
\end{equation}
Then, the tau-function for JM Lax pair is defined by:
\begin{equation} 
\label{impor} -\hbar^2 \frac{d}{dt} \ln \tau_{\text{JM}}= \sigma(t).
\end{equation}

Since \eqref{eqtau} only involves even power of $\hbar$ then the $\hbar$-expansion of $\sigma(t)$ and $\ln\tau(t)$ only involve even powers of $\hbar$:
\begin{equation} 
\label{tauexp}\sigma(t)=\sum_{k=0}^\infty \sigma_{2k}(t) \hbar^{2k}, 
\end{equation}
\begin{eqnarray} \label{zexp} 
\ln \tau_{\rm JM} &\overset{\text{def}}{=}&
  \sum_{k=0}^\infty \tau_{2k}(t) \hbar^{2k-2} , \quad 
\tau_{2k}(t) = -\int^t \sigma_{2k}(s)ds. 
\end{eqnarray}
Moreover, \eqref{relation} implies that $p(t)$ also only contains even order terms: 
\begin{equation}  \label{pexp} 
p(t) = \sum_{k=0}^\infty p_{2k}(t) \hbar^{2k}. 
\end{equation}

\subsection{\label{sec2}First orders of the JM tau-function}

In this section we present the computation of the first orders of the tau-function for the JM Lax pair. In what follows, we choose to express all quantities as function of $q_0(t)$ which is a solution of \eqref{y0JM}. 
Straightforward computations give: 
\begin{eqnarray}  
\sigma_0(t)&=&\frac{\theta(8q_0^3-\theta)}{8q_0^2},\cr
\sigma_2(t)&=&-\frac{\theta q_0}{8(4q_0^3+\theta)^2},\cr
\sigma_4(t)&=&-\frac{3\theta q_0^4(560q_0^6-184\, \theta q_0^3+3\theta^2)}{32(4q_0^3+\theta)^7},\cr
\sigma_6(t)&=& -\frac{\theta q_0^7}{32\left(4q_0^3+\theta\right)^{12}}
\bigl( 3203200\,q_0^{12}-3668064\,\theta q_0^9 
\cr &   & 
+838632\,\theta^2q_0^6 - 39482\,\theta^3q_0^3+189\,\theta^4 \bigr). 
\label{sigm} 
\end{eqnarray}
In particular one can also verify directly that the coefficients presented here satisfy the differential equation \eqref{eqtau}. 
Integrating over $t$ leads to:
\begin{eqnarray} \label{taunum} 
\tau_0(t)&=&\frac{4}{3}\theta q_0^3+
\frac{\theta^3}{24q_0^3}+\frac{\theta^2}{2}\ln q_0 +\text{Cste},\cr
\tau_2(t)&=&\frac{1}{24}\ln(1+\frac{\theta}{4q_0^3})+\frac{1}{12}
\ln 2 +\text{Cste},\cr
\tau_4(t)&=&\frac{\theta(700\,q_0^6-85\, \theta q_0^3-2\theta^2)}
{480(4q_0^3+\theta)^5}+\text{Cste},\cr
\tau_6(t)&=&\frac{\theta}{4032\left(4q_0^3+\theta\right)^{10}}
\bigl(
6726720\, q_0^{15}-5017712\,\theta q_0^{12}
 \nonumber \\[+.3em] 
& & 
+541132\,\theta^2 q_0^9
-1089\,\theta^3q_0^6
+160\,\theta^4q_0^3+4\,\theta^5
\bigr) + \text{Cste}  \hspace{+1.5em}
\end{eqnarray}
Here the constant terms are to be understood as not depending on $t$. 
We will see that these integration constants are specified by the topological recursion, and correspond to lower end-points for the integral \eqref{zexp} defining $\tau_{2k}$ taken at $t = \infty$ for $k\geq 2$. Actually, we can choose $\infty$ (for any $\infty_A, \infty_B$ and $\infty_C$) as the lower end-point since $\sigma_{2k} \underset{t\to\infty}{=} O(t^{-2})$ holds if $k \ge 2$. This property can easily be proved by a recursion relation satisfied by $\{ q_{i}(t) \}_{i \ge 0}$ appearing in \eqref{eq:formal-solution-q}.

\subsection{Spectral curve and topological recursion for the JM pair} 
From Berg\`ere and Eynard \cite{Deter} we know that the spectral curve of a Lax pair is given by the leading order in $\hbar$ of the characteristic polynomial of $\mathcal{D}(x,t)$. Thus we find:
\begin{eqnarray} \label{SpecCurveJM}
y^2 & = & (x-q_0)^2(x^2+2q_0x+3q_0^2+t) \nonumber \\[+.2em]
& = & (x-q_0)^2\left(x+q_0-\sqrt{-\frac{\theta}{q_0}}\right)
\left(x+q_0+\sqrt{-\frac{\theta}{q_0}}\right)
\end{eqnarray}
where $q_0(t)$ is the solution of \eqref{y0JM}. 
We call \eqref{SpecCurveJM} the Jimbo-Miwa spectral curve (JM curve, for short). JM curve is of genus $0$ with two branch points. 
It can be parametrized with a global Zhukovsky variable:
\begin{eqnarray} \label{SpecCurveJM2} \left\{
\begin{array}{l}
\displaystyle 
x(z)=-q_0+\frac{1}{2}\sqrt{-\frac{\theta}{q_0}}
\left(z+\frac{1}{z}\right), \\
\displaystyle 
y(z)=\frac{1}{2}\sqrt{-\frac{\theta}{q_0}}\left(z-\frac{1}{z}\right)\left(-2q_0+\frac{1}{2}\sqrt{-\frac{\theta}{q_0}}\left(z+\frac{1}{z}\right)\right).\\
\end{array}
\right.\end{eqnarray}
With this parametrization, the branch points are located at $z=\pm 1$ and the differential $ydx$ has two poles at $z=0$ and $z=\infty$. 

\begin{defn}[Definition 4.2 of \cite{EO}]
For $g \ge 0$ and $n \ge 1$, define the Eynard-Orantin differential 
$\omega^{(g)}_{n}(z_{1},\dots,z_{n})$ of the type $(g,n)$ for the spectral curve \eqref{SpecCurveJM2} 
by the following topological recursion relation (\cite{EO}):
\begin{eqnarray}
\omega^{(0)}_{1}(z_{1}) & = & y(z_{1})dx(z_{1}), \cr
\omega^{(0)}_{2}(z_{1},z_{2}) & = & 
\frac{dz_{1}dz_{2}}{(z_{1}-z_{2})^{2}}, \cr 
\omega^{(g)}_{n+1}(z_{0},z_{1},\dots,z_{n}) & = & 
\sum_{r\in R} \Res_{z=r} \,K(z_{0},z) \Bigl[ \omega^{(g-1)}_{n+1}(z,\bar{z},z_{1},\dots,z_{n}) \cr 
& & + \sum'_{\substack{g_{1} + g_{2} = g \cr 
I \cup J = \{1,\dots,n\} }} \omega^{(g_{1})}_{1+|I|}(z,z_{I})\omega^{(g_{2})}_{1+|J|}(\bar{z},z_{J}) \Bigr]. \hspace{+2.5em}\label{eq:top-rec} 
\end{eqnarray}
Here $R$ is the set of branch points, $\bar{z}$ is the local conjugate of $z$ near a branch point $r$, 
\begin{equation}
K(z_{0},z) = \frac{1}{2}\frac{\int^{\bar{z}}_{z} 
\omega^{(0)}_{2}(\cdot, z_{0})}{(y(z)-y(\bar{z}))dx(z)}
\end{equation}
is called the recursion kernel, and the summation in the last line of \eqref{eq:top-rec} means ``except for the cases $(g_{1}, I) = (0, \emptyset)$ and $(g_{2}, J) = (0, \emptyset)$''. 
\end{defn}

For JM curve \eqref{SpecCurveJM2}, $R = \{+1,-1 \}$ and $\bar{z} = z^{-1}$ for both branch points $r = \pm 1$. The Eynard-Orantin differentials $\omega^{(g)}_{n}$ are meromorphic multi-differentials on the $n$-times product of spectral curves, and known to be holomorphic except for the branch points if $(g,n) \ne (0,1), (0,2)$. Since our spectral curve has genus 0, the topological recursion becomes easier (see \cite{EO} for general case). 

We also introduce symplectic invariants $F^{(g)}$ for a genus $0$ spectral curve, following \cite{EO}: 

\begin{defn}[Definition 4.3 of \cite{EO}] 
The $g$-th symplectic invariant (or genus $g$ free energy) of the spectral curve is defined as follows:
\begin{itemize}
\item 
For $g = 0$, set 
\begin{equation} \label{eq:def-symplectic-invariant-0}
F^{(0)} = \frac{1}{2}\sum_{\alpha} \Res_{z=\alpha} V_{\alpha}(z)y(z)dx(z) 
+ \frac{1}{2} \sum_{\alpha} t_{\alpha} \mu_{\alpha},
\end{equation}
where the sum is taken over all poles of $y(z)dx(z)$, 
and for a pole $\alpha$ of $y(z)dx(z)$, we define 
\begin{eqnarray}
t_{\alpha} & = & \Res_{z=\alpha} y(z)dx(z) \\
V_{\alpha}(z) & = & \Res_{p = \alpha} ~
\ln \Bigl( 1 - \frac{\xi_{\alpha}(z)}{\xi_{\alpha}(p)} \Bigr) y(p) dx(p), \\
\mu_{\alpha} & = & 
\int^{z_o}_{\alpha} \Bigl( y(z)dx(z) - dV_{\alpha}(z) 
- t_{\alpha} \dfrac{d\xi_\alpha(z)}{\xi_{\alpha}(z)} \Bigr)  \nonumber \\
& & + V_{\alpha}(z_o) + t_{\alpha} \ln \xi_{\alpha}(z_o).
\end{eqnarray}
Here $z_o$ is a fixed arbitrary generic point on the spectral curve, 
and $\xi_{\alpha}(z)$ is a certain function of $z$ which is chosen 
depending on the behavior of $y(z)dx(z)$ near $z = \alpha$
(see \S 3.4.2 of \cite{EO}).  

\smallskip
\item 
For $g = 1$, set
\begin{equation} \label{eq:def-symplectic-invariant-1}
F^{(1)} = - \frac{1}{24} \ln \left( 
\tau_{B}(\{ x(a_i) \})^{12} \prod_{r \in R} y'(r) \right),
\end{equation}
where $\tau_{B}(\{ x_i \})$ is a function of $x_i = x(r_i)$ ($r_i \in R$) 
which is defined (up to constant) by the following property:
\[
\dfrac{\partial}{\partial x_i} \ln \tau_{B}(\{x_i \}) 
= \Res_{z=a_i} \frac{B(z,\bar{z})}{dx(z)}.
\]
($\tau_{B}(\{ x_i \})$ is called the Bergman tau-function of the spectra curve.)
We also set 
\begin{equation} \label{eq:y-prime-r}
y'(r) = \lim_{z \to r} \frac{y(z) - y(r)}{\sqrt{x(z) - x(r)}}
\end{equation}
for $r \in R$.

\smallskip
\item 
For $g \ge 2$, set
\begin{equation} \label{eq:def-symplectic-invariant}
F^{(g)} = \frac{1}{2-2g}
\sum_{r \in R} \Res_{z=r} \Phi(z) \omega^{(g)}_{1}(z),
\end{equation}
where 
\[
\Phi(z) = \int^{z}_{z_{o}} y(z)dx(z)
\]
with an arbitrary chosen generic point $z_o$.
\end{itemize}
\end{defn}

Denote by $F^{(g)}_{\rm JM}$ the symplectic invariants 
for JM curve \eqref{SpecCurveJM2}. 
In the above definition of $F^{(0)}_{\rm JM}$, 
$y(z)dx(z)$ has poles at $0$ and $\infty$, 
and we chose $\xi_{0}(z) = \xi_{\infty}(z) = x(z)$. 
Also, for $F^{(1)}$, we fix the ambiguity of 
the normalization constant of 
the Bergman tau-function as 
\[
F^{(1)}_{\rm JM} = - \frac{1}{24} \ln \left( 
\gamma^3 y'(1) y'(-1)
\right), \quad \gamma = \frac{1}{2} \sqrt{-\frac{\theta}{q_0}}.
\]
(See Chapter 7 of \cite{E-book} for the above formula.)
Then, we find:
\begin{eqnarray} \label{res1}
F_{\rm JM}^{(0)}&=&\frac{4\theta}{3}q_0^3-\frac{\theta^2}{4}+\frac{\theta^3}{24q_0^3}
- \frac{\theta^2}{2} \ln \left(- \frac{\theta}{4q_0} \right)
\cr
F_{\rm JM}^{(1)}&=&-\frac{1}{24}\ln\left(\theta^2\left(1+\frac{\theta}{4q_0^3}\right)\right),\cr
F_{\rm JM}^{(2)}&=&{\frac { \left( 
2048\,{q_0}^{12}+2560\,\theta\,{q_0}^{9}+1280\,{\theta}^{2}{q_0}^{6}
+1020\,{\theta}^{3}{q_0}^{3}-45\,{\theta}^{4} 
\right) {q_0}^{3}}
{480\, {\theta}^{2}\left( 4\,{q_0}^{3}+\theta \right) ^{5}}},  \cr
F_{\rm JM}^{(3)}&=&-\frac {q_0^6}{4032\theta^4 \left( 
\theta+4\,q_0^3 \right) ^{10}}\, 
\Big(4194304\,q_0^{24}+10485760\,\theta\,q_0^{21}  \cr 
& & 
+11796480\,\theta^2q_0^{18} +7864320\,\theta^3q_0^{15}+3440640\,\theta^4q_0^{12} \cr 
&& -5694528\,\theta^5q_0^9+5232752\,\theta^6q_0^6
-510412\,\theta^7q_0^3+3969\,\theta^8 \Big). \hspace{+1.5em}  
\end{eqnarray}
Moreover, it is easy to prove that when $q_0\to 0$ (i.e., $t\to \infty_A$) the correlation functions $\omega_n^{(g)}$ and the symplectic invariants $F_{\rm JM}^{(g)}$ (identified with $\omega_0^{(g)}$ in the next formula) behave like:
\begin{equation}  \omega_n^{(g)}(z_1,\dots,z_n) \underset{q_0\to 0}{\sim}\text{Cste}\, q_0^{\frac{3}{2}(2g-2+n)} dz_{1}\cdots dz_{n}.
\end{equation}
Indeed, the kernel $K(z_0,z)$ used in the recursion behaves like:
\begin{equation} K(z_0,z)=-\frac{4z^4}{(z^2-1)(z_0z-1)(z_0-z)
\left( (1+z)\left( 
\frac{-\theta}{q_0}\right)^{\frac{3}{2}}+4\theta z \right)}\frac{dz_{0}}{dz} =O\left(q_0^{\,\frac{3}{2}}\right). \end{equation}
Thus adding a power $q_0^{3/2}$ at each step of the recursion. In particular, we get:
\begin{equation} \label{eq:FgJM-infty-A}
\underset{t\to \infty_A}{\lim} F_{\rm JM}^{(g)}(t)=0 \quad \text{for $g \ge 2$}.\end{equation}

\subsection{Tau-function and symplectic invariants for the JM pair}

In this subsection we state one of our main results regarding the relationship between symplectic invariants $F^{(g)}_{\rm JM}$ and the tau-function of Painlev\'e 2. 

\begin{thm}\label{propo1} 
The JM pair is of topological type (in the sense of Section \ref{DeterForm}) and we have: 
\beq \label{eq:fg-and-tau-JM-theorem}
\frac{dF_{\rm JM}^{(g)}(t)}{dt} = - \sigma_{2g}(t) \quad 
\text{for $g \ge 2$}. \eeq
\end{thm}

Theorem \ref{propo1} will be proved in Section \ref{DeterForm} and in appendix. From former results \eqref{sigm} and \eqref{res1} we can verify that $\frac{dF_{\rm JM}^{(g)}}{dt} = - \sigma_{2g}$  also hold for $g = 0$ and $g = 1$. This implies the following:

\begin{thm} \label{thm:JMFg-tau}
The generating function of symplectic invariants of JM curve \eqref{SpecCurveJM2} gives a tau-function of Painlev\'e $2$. In other words,  
\beq
\ln \tau_{\text{\rm JM}} = \sum_{g=0}^{\infty} \hbar^{2g-2} F^{(g)}_{\rm JM}(t) 
\eeq
satisfies \eqref{impor}. Furthermore, we have
\begin{equation} \label{eq:FgJM-and-inftyA}
F_{\rm JM}^{(g)}(t)= - \int_{\infty_A}^t \sigma_{2g}(s) ds \quad \text{for $g \ge 2$}.
\end{equation}
\end{thm}
Theorem \ref{thm:JMFg-tau} follows from Theorem \ref{propo1} and \eqref{eq:FgJM-infty-A}. 

\subsection{Limit at $\infty_{B}$ and $\infty_{C}$: The Hermite-Weber curve} \label{section:Limit-JM-to-Weber}
We already know that  the functions $F^{(g)}_{\rm JM}(t)$ vanish for $g\geq 2$ when $t \to \infty_A$ (i.e., $q_0\to 0$). We find that some interesting numbers appear when taking the limit of $F^{(g)}_{\rm JM}(t)$ to both $t \to \infty_{B}$ and $t \to \infty_{C}$. Unfortunately taking these limits $t \to \infty_{B,C}$ in the spectral curve \eqref{SpecCurveJM2} is not directly possible since the coefficients diverge in the limit. To avoid this difficulty, we perform the following affine symplectic transformation
\beq \label{eq:coord-tr-JM} x=\frac{1}{2\sqrt{-q_0}}X-q_0\,\,\,,\,\,\, y=2\sqrt{-q_0}\, Y\eeq
giving the curve:
\beq \label{eq:JM-curve-before-limit}
Y^2=\left(\frac{X^2}{4}-\theta\right)\left(1+\frac{X}{4(-q_0)^{\frac{3}{2}}}\right)^{2}. 
\eeq
It is clear that the spectral curve \eqref{eq:JM-curve-before-limit} has the same symplectic invariants $F^{(g)}_{\rm JM}$ since the coordinate transformation \eqref{eq:coord-tr-JM} preserves the 1-form $\omega^{(0)}_1$ and the set of branch points. In the limit $t\to \infty_{B,C}$ (i.e., $q_{0} \rightarrow \infty$), the curve \eqref{eq:JM-curve-before-limit} is reduced to the Hermite-Weber curve: 
\beq Y^2=\frac{X^2}{4}-\theta\eeq
which can be parametrized into:
\beq \label{SpecCurveWeber} \left\{
\begin{array}{l}
 \displaystyle 
X(z)=\sqrt{\theta}\left(z+\frac{1}{z}\right), \\
 \displaystyle
Y(z)=\frac{\sqrt{\theta}}{2}\left(z-\frac{1}{z}\right).\\
\end{array}
\right.\eeq

\begin{prop}[Cf. \cite{HZ, Penner}] \label{prop:Fg-Weber-curve}
The symplectic invariants $F^{(g)}_{\rm Weber}$ for the spectral curve \eqref{SpecCurveWeber} are given by 
\bea 
\label{eq:Weber-F0}
F^{(0)}_{{\rm Weber}}&=&
\frac{3\theta^2}{4}-\frac{\theta^2}{2}\ln \theta, \\
\label{eq:Weber-F1}
F^{(1)}_{{\rm Weber}}&=&-\frac{1}{12}\ln \theta,\\
\label{eq:Weber-Fg}
F^{(g)}_{{\rm Weber}}&=&-\frac{B_{2g}}{2g(2g-2)\theta^{2g-2}} \quad \text{for $g \ge 2$}. 
\eea
\end{prop}
\begin{proof}
Although the statement is well-known in the physics literature (see Remark \ref{remark:Gaussian-free-energy} below), we could not find a specific proof in the literature. Therefore we propose here an alternative proof based on the results obtained in \cite{Dumitrescu-Mulase, Mulase-Penkava}. In order to use these results, let us consider the spectral curve: 
\beq \label{SpecCurveWeber-2} \left\{
\begin{array}{l}
 \displaystyle 
\tilde{X}(t)=\sqrt{\theta}\left(2+\frac{4}{t^2 - 1}\right), \\
 \displaystyle
\tilde{Y}(t)=\sqrt{\theta}\, \frac{t+1}{t-1}.\\
\end{array}
\right.\eeq
It is obtained from the spectral curve \eqref{SpecCurveWeber} by the affine symplectic transformation $(X,Y) \mapsto (\tilde{X}, \tilde{Y}) = \left(X, Y + \frac{X}{2}\right)$ which preserves the set $\left(F^{(g)}\right)_{g\geq 2}$ (Theorem $7.1$ of \cite{EO}), combined with the change of parametrization $z \mapsto z(t) = \frac{t+1}{t-1}$. In \cite{Mulase-Penkava} the authors introduced the functions (called the Poincar\'e polynomials): 
\[
F^{(g)}_{n}(t_1,\dots, t_n) = \left(
\sum_{\substack{\Gamma:~\text{ribbon graphs} \\ \text{of type}~(g,n)}}
\frac{(-1)^{e(\Gamma)}}{|{\rm Aut}(\Gamma)|}
\prod_{\substack{\eta:~\text{edges} \\ \text{in}~\Gamma}}
\frac{(t_{i_{\eta}}+1) (t_{j_{\eta}} + 1)}
{2(t_{i_{\eta}}+t_{j_{\eta}})} \right) \theta^{2-2g-n}
\]
for $g \ge 0$ and $n \ge 1$ (see \cite{Mulase-Penkava} for the precise definitions of the involved quantities). In \cite{Dumitrescu-Mulase, Mulase-Penkava}, the authors proved that these functions satisfy the following properties: 
\begin{itemize}
\item 
For $g \ge 0$ and $n \ge 1$, the multi-differentials 
\[
\tilde{\omega}^{(g)}_n(t_1,\dots, t_n) = 
d_{t_1}\cdots d_{t_n}F^{(g)}_{n}(t_1,\dots,t_n)
\] 
satisfy the topological recursion applied to the spectral curve \eqref{SpecCurveWeber-2} (See Theorem $2.13$ of \cite{Dumitrescu-Mulase}). 
\item For $g \ge 0$ and $n \ge 1$, we have $F^{(g)}_{n}(t_1,\dots,t_n)|_{t_i=-1} = 0$ for each $1\leq i \leq n$.
\item For $g \ge 0$ and $n \ge 1$, we have $F^{(g)}_{n}(1,\dots,1) = (-1)^n \chi({\mathcal M}_{g,n}) \theta^{2-2g-n}$, where ${\mathcal M}_{g,n}$ is the moduli space of genus $g$ Riemann surfaces with $n$ marked points. In particular, it is proved in \cite{HZ, Penner} that:
\begin{equation} \label{eq:formula-for-moduli-space}
\chi({\mathcal M}_{g,n}) = 
\frac{(2g-3+n)!}{(2g)!} (2g-1) B_{2g}.
\end{equation}
\end{itemize}
Precisely speaking, \cite{Dumitrescu-Mulase, Mulase-Penkava} considered the case $\theta = 1$. But these results can be generalized as above since we can introduce the parameter $\theta$ by a trivial scaling of variables. In particular, we have 
$\int^{t}_{-1} \tilde{\omega}^{(g)}_{1}(t) = F^{(g)}_1(t)$
and thus 
\[
\int^{1}_{-1} \tilde{\omega}^{(g)}_{1}(t) 
= F^{(g)}_1(1)
= - \chi({\mathcal M}_{g,1}) \theta^{1-2g} 
= \frac{B_{2g}}{2g \, \theta^{2g-1}}.
\]
On the other hand, since the parameter $\theta$ is the filling fraction (or equivalently $\pm \theta$ are the temperatures at $t = \pm 1$, respectively), the variation formula (see Section $5.3$ of \cite{EO}) implies that: 
\[ 
\frac{d F^{(g)}_{\rm Weber}}{d \theta} 
= \int^{1}_{-1} \tilde{\omega}^{(g)}_{1}(t) 
= \frac{B_{2g}}{2g \, \theta^{2g-1}} \quad 
\text{for $g\geq 2$}.
\]
Thus we have the desired equality \eqref{eq:Weber-Fg} modulo an additive constant which is independent of $\theta$. However, this additive constant must vanish due to the homogeneity relation  
\[ 
F^{(g)}_{\rm Weber}\bigl|_{\theta \mapsto \lambda \theta} 
= \lambda^{2-2g} F^{(g)}_{\rm Weber} 
\quad \text{for $g\geq 2$}.
\]
(See Theorem $5.3$ of \cite{EO}).
Thus, we have finally proved \eqref{eq:Weber-Fg} for $g\geq 2$. Eventually, formulas \eqref{eq:Weber-F0} and \eqref{eq:Weber-F1} are obtained from a direct explicit computation.
\end{proof}

\begin{rem} \label{remark:Gaussian-free-energy}
The curve \eqref{SpecCurveWeber} also appears as the spectral curve arising in the following Gaussian Hermitian matrix model (with an appropriate normalization): 
\beq \label{DefMatrix}
Z_{\rm G} =\frac{1}{(2\pi)^N N!}\int_{\mathbb{R}^{N}}d\lambda_1\dots d\lambda_{N} \,\Delta(\lambda_1,\dots,\lambda_{N})^2\, e^{-\frac{1}{\hbar} \underset{i=1}{\overset{N}{\sum}}\frac{\lambda_i^2}{2}}. 
\eeq
Here 
$\Delta(\lambda_1,\dots,\lambda_{N}) = \underset{1\leq i<j\leq N}{\prod}  (\lambda_j - \lambda_i)$ 
is the standard Vandermonde determinant. 
The value of the partition function $Z_{\rm G}$ is explicitly known from Mehta's integral (a case of Selberg-like integrals, see \cite{BGOneCut} for details):
\begin{equation} \label{MehtaIntegrals}
Z_{\rm G} = \frac{\hbar^{\frac{N^2}{2}}}{(2\pi)^{\frac{N}{2}} N!}\,
\underset{i=1}{\overset{N}{\prod}}\Gamma(1+i)=\frac{\hbar^{\frac{N^2}{2}}}{(2\pi)^{\frac{N}{2}}}\,
\underset{i=1}{\overset{N}{\prod}}i!.
\end{equation}
The parameter $\theta$ in \eqref{SpecCurveWeber} is the so-called 't Hooft parameter in the analysis of the large N limit: 
$N \to \infty$, $\hbar \to 0$ with $\theta = N \hbar$ fixed. 
The above explicit formula enables us to find the explicit large $N$ expansion (see \cite{Marino} for example):
\[
- \ln Z_{G}\bigl|_{N = \hbar \theta} ~\sim~ 
\sum_{g \ge 0} \hbar^{2g-2} F^{(g)}_{\rm G}(\theta),
\]
\[
F^{(0)}_{\rm G}(\theta) = \frac{3}{4}\theta^2 - \frac{1}{2}\theta^2 \ln \theta, 
\quad
F^{(1)}_{\rm G}(\theta) = - \frac{1}{12} \ln \theta + \zeta'(-1), 
\]
\begin{equation} \label{eq:Gaussian-free-energy}
F^{(g)}_{\rm G}(\theta) = - \frac{B_{2g}}{2g(2g-2) \theta^{2g-2}} 
\quad \text{for $g \ge 2$}.
\end{equation}
Here $\zeta'(-1)$ is the derivative of Riemann's $\zeta$-function $\zeta(s)$ evaluated at $s=-1$. The right hand side of \eqref{eq:Gaussian-free-energy} coincides with the Euler characteristic of the moduli space ${\mathcal M}_g$ of genus $g$ Riemann surfaces (see \cite{HZ, Penner}). 

For $\theta=1$, general results regarding the connection of the topological recursion with standard large $N$ limit asymptotic expansions of Hermitian matrix integrals (with polynomial potentials) claim that the generating function of the symplectic invariants $F^{(g)}_{\rm Weber}\bigl|_{\theta=1}$ matches with the asymptotic expansion of $-\ln Z_G\bigl|_{\theta=1}$ (which is the standard large $N$ limit of the Gaussian Hermitian matrix integral) up to some additive constants: (See Corollary $5.1$ in \cite{Chekhov-Eynard-Oranrin} or main theorems of \cite{BBEnew,BGOneCut}). More precisely, we have:  
\[
F_{\rm Weber}^{(g)}\bigl|_{\theta=1} = 
F_{G}^{(g)}\bigl|_{\theta=1} + C^{(g)} \quad \text{for $g\geq 0$}.
\]
Since $\theta$ may be introduced by a trivial rescaling of the parameter $N\mapsto N\theta$ in the Gaussian matrix integral, the last equality may be extended to:
\beq F_{\rm Weber}^{(g)}(\theta) = F_{G}^{(g)}(\theta) + C^{(g)}\theta^{2-2g} \quad \text{for $g\geq 0$}. \eeq  
Using the exact expression \eqref{eq:Gaussian-free-energy} and the values of $F_{\text{Weber}}^{(g)}(\theta)$ proved in Proposition \ref{prop:Fg-Weber-curve}, we may obtain the constants and we find:
\[ 
C^{(0)}=0 \,,\, C^{(1)}=\zeta'(-1) \text{ and } 
C^{(g)}=0 \text{ for } g\geq 2. 
\]
Note in particular that, as claimed in the physics literature, the constants $\left(C^{(g)}\right)_{g\geq 2}$ are vanishing. However, we stress here that current theorems relating the topological recursion with Hermitian matrix integrals are not sufficient to determine the constants and thus prove Proposition \ref{prop:Fg-Weber-curve}.

Eventually, we note that the symplectic invariants \eqref{eq:Weber-Fg} are closely related to the Voros coefficients in the theory of the exact WKB analysis. This issue will be discussed in the forthcoming paper \cite{IKT}.
\end{rem}

Since it is known from \cite{EO} that the symplectic invariants (and correlation functions) obtained for a limiting curve are equal to the limit of symplectic invariants, we have from proposition \ref{prop:Fg-Weber-curve} the following theorem: 
\begin{thm}\label{propo3}
Both of the limits $t\to \infty_{B}$ and $t \to \infty_{C}$ of
the symplectic invariant $F^{(g)}_{\rm JM}(t)$ of JM curve are given by
\beq \label{ttt}
\underset{t\to \infty_{B,C}}{\lim} F^{(g)}_{\rm JM}(t)=-\frac{B_{2g}}{2g(2g-2)\theta^{2g-2}} \quad \text{for $g \ge 2$}. \eeq
\end{thm}

In particular taking $q_0 \rightarrow 0$ in \eqref{res1} we can verify that this result holds for $g=2$ and $g=3$. Theorem \ref{propo3} and equation \eqref{eq:FgJM-and-inftyA} also imply the following:
\begin{equation}\label{eq:int-sigma-Bernoulli}
\int^{\infty_{B,C}}_{\infty_A} \sigma_{2g}(t)dt = \frac{B_{2g}}{2g(2g-2)\theta^{2g-2}} \quad \text{for $g \ge 2$}.
\end{equation}
Note that a path connecting $\infty_A$ and $\infty_{B, C}$ never exists when $\theta = 0$ since the equation \eqref{y0JM} defining $q_0(t)$ splits in that case. This is consistent with the fact that the r.h.s. of \eqref{eq:int-sigma-Bernoulli} blows up when $\theta\to 0$.


\section{The Harnad-Tracy-Widom Lax Pair}

In this section, we develop the same approach for the Painlev\'e $2$ system but with another Lax pair, called Harnad-Tracy-Widom Lax pair, as a starting point. This Lax pair is connected to the previous one \eqref{JMLaxPair} by a Laplace-type integral transformation (\cite{JKT}), but their relation is non-trivial in terms of the topological recursion. To our knowledge the two Lax pairs (up to trivial transformations) studied in this article represent the two usual pairs used to describe the Painlev\'e $2$ system. 

\subsection{Introduction of $\hbar$ in the Harnad-Tracy-Widom Lax pair}

The following Lax pair is introduced in \cite{HTW}: 
\bea \label{eq:original-HTW}
\left\{\begin{array}{lll} 
\displaystyle
\frac{\partial }{\partial x} \Psi(x,t)&=&
\begin{pmatrix} 
\displaystyle -q+\frac{\theta}{2x}&
\displaystyle x-p-2q^2-t\\
\displaystyle \frac{1}{2}+\frac{p}{2x}&
\displaystyle q-\frac{\theta}{2x} \end{pmatrix}\Psi(x,t),\\[+1.9em]
\displaystyle \frac{\partial }{\partial t} \Psi(x,t)&=&
\begin{pmatrix} q&-x\\
\displaystyle -\frac{1}{2}& -q\end{pmatrix}\Psi(x,t). \cr
\end{array}
\right.\eea
Like in the JM case we can introduce a small expansion parameter $\hbar$ with a suitable rescaling of the variables:
\beq 
x\to \hbar^{-\frac{2}{3}}\td{x}\,\,,\,\, q\to \hbar^{-\frac{1}{3}}\td{q}\,\,,\,\, p\to \hbar^{-\frac{2}{3}}\td{p}\,\,,\,\,t\to \hbar^{-\frac{2}{3}}\td{t}\,\,,\,\, \theta\to \hbar^{-1}\td{\theta}
\eeq
as well as a suitable gauge transformation $\Psi \to {\rm diag}(\hbar^{-\frac{1}{6}}, \hbar^{\frac{1}{6}}) \Psi$. Omitting $\tilde{~}$ for clarity, we get: 
\beq  \label{HTWLaxPair}
\left\{
\begin{array}{lll} 
\displaystyle 
\hbar\frac{\partial }{\partial x} \Psi(x,t)&=&
\begin{pmatrix} \displaystyle  -q+\frac{\theta}{2x}&
\displaystyle  x-p-2q^2-t\\
\displaystyle  \frac{1}{2}+\frac{p}{2x}&
\displaystyle  q-\frac{\theta}{2x}\end{pmatrix}\Psi(x,t) \cr
& \hs{-.1em} \overset{\text{def}}{=} & 
\mathcal{D}(x,t) \Psi(x,t),  \\[+.5em]
\displaystyle  
\hbar\frac{\partial }{\partial t} \Psi(x,t)&=&
\begin{pmatrix} q&-x\\
\displaystyle -\frac{1}{2}& -q\end{pmatrix}\Psi(x,t) \cr 
&  \hs{-.1em} \overset{\text{def}}{=} & 
\mathcal{R}(x,t) \Psi(x,t).\cr
\end{array}
\right.\eeq
We call the Lax pair \eqref{HTWLaxPair} Harnad-Tracy-Widom pair (HTW pair, for short).

The compatibility equations for this Lax pair are given by:
\beq \label{Eq}\hbar\dot{q}=q^2+p+\frac{t}{2}\,\,\,\text{ and }\,\,\, \hbar\dot{p}=-2qp-\theta. \eeq
As in the Jimbo-Miwa case, we recover that $q$ is a solution of the Painlev\'e $2$ equation:
\beq \hbar^2\ddot{q}=2q^3+tq-\theta+\frac{\hbar}{2}.\eeq
Although the compatibility equations are the same as for the JM pair, the definition of Jimbo-Miwa-Ueno tau-function is a little different (\cite{JMI,JMII}; see also Section \ref{subsection:TTP}). The Hamiltonian for the HTW pair is: 
\begin{equation}
H_{\rm HTW} = \frac{1}{2}p^2 + \left(q^2 + \frac{t}{2} \right) p + \theta q + \frac{t^2}{8}
\end{equation}
and the tau-function for HTW pair is defined as 
\begin{equation} \label{impor-HTW}
-h^2 \frac{d}{dt}\,{\rm ln}\,\tau_{\text{HTW}} = H_{\rm HTW} = \sigma(t) + \frac{t^2}{8}
\end{equation}
where $\sigma(t)$ is the solution of the $\sigma$-form of Painlev\'{e} $2$ in \eqref{eqtau}. 

\subsection{Spectral curve and topological recursion for the HTW pair}
As usual the spectral curve is given by the leading order in $\hbar$ of the characteristic polynomial of $\mathcal{D}(x,t)$. We find:
\beq \label{SpecCurveHTW} y^2=\frac{1}{2x^2}\left(x-\frac{\theta}{2q_0}\right)^2\left(x+2q_0^2\right)\eeq
where $q_0(t)$ is a solution of \eqref{y0JM}. It is a genus $0$ curve with a single branch point arising at $x=-2q_0^2$ but with a pole singularity at $x=0$. It can be parametrized globally with:
\beq \label{SpecCurveHTW2} \left\{
\begin{array}{l}
 \displaystyle x(z)=2q_0^2(z^2-1), \\[+.2em]
 \displaystyle y(z)=\frac{z\left( \theta - 4 q_0^3(z^2-1)\right)}{4q_0^2\left(z^2-1\right)}=\frac{z\left(q_0 x(z)-
  \frac{\theta}{2}\right)}{x(z)}.\\
\end{array}
\right.\eeq
We call this spectral curve the HTW spectral curve. Note that the branch point is located at $z=0$ and the local conjugate point around the branch point is given by $\bar{z}=-z$. We define the Eynard-Orantin differentials and the symplectic invariants $F^{(g)}_{\rm HTW}$ for the HTW spectral curve \eqref{SpecCurveHTW2} in the same way as \eqref{eq:top-rec}, \eqref{eq:def-symplectic-invariant-0}, \eqref{eq:def-symplectic-invariant-1} and \eqref{eq:def-symplectic-invariant}.  
To be more precise, in the definition of $F^{(0)}_{\rm HTW}$, 
the sum in \eqref{eq:def-symplectic-invariant-0} runs over 
$\alpha \in \{+1, -1, \infty \}$ and 
$\xi_{1}(z) = \xi_{-1}(z) = x(z)^{-1}$ and $\xi_{\infty}(z) = x(z)^{1/2}$.
On the other hand, following Chapter 7 of \cite{E-book}, 
we define $F^{(1)}_{\rm HTW}$ by 
\[
F^{(1)}_{\rm HTW} = - \frac{1}{24} \ln \left( y'(0) \right).
\]
Here $y'(0)$ is defined in \eqref{eq:y-prime-r}. Then we find:
\bea \label{eq:free-energy-HTW}
F_{\rm HTW}^{(0)}(t)&=&
\frac{q_0^6}{3}+\frac{5\theta}{6} q_0^3 -\frac{3\theta^2}{4}
+\frac{\theta^2}{4} \ln \left( - 8q_0^2 \right)
\nonumber \\[+.2em]
F_{\rm HTW}^{(1)}(t)&=& -\frac{1}{24}\ln\left( \frac{1}{\sqrt{2}} 
\Bigl( 1+\frac{\theta}{4q_0^3}\Bigr) \right), \nonumber \\[+.2em]
F_{\rm HTW}^{(2)}(t)&=&\frac{\theta\left(700\,q_0^6-85\,\theta\,q_0^3-2\theta^2\right)}
{480\left(4q_0^3+\theta\right)^5}, \nonumber \\[+.3em]
F_{\rm HTW}^{(3)}(t)&=&
\frac{\theta}{4032 \left(4q_0^3+\theta\right)^{10}}
\bigl(6726720 q_0^{15}
-5017712\theta q_0^{12} \nonumber \\[+.3em]
& & 
+541132\theta^2 q_0^9
-1089\theta^3q_0^6
+160\theta^4q_0^3
+ 4\theta^5  \bigr). 
\eea
Additionally, as $q_0\to \infty$ (i.e., $t\to \infty_{B,C}$), it is easy to prove by recursion that the correlation functions and symplectic invariants $F^{(g)}_{\rm HTW}$ (identified with $\omega_0^{(g)}$ in the next formula) generated by the topological recursion on \eqref{SpecCurveHTW2} behave like:
\beq  \omega_n^{(g)}(z_1,\dots,z_n) \underset{q_0\to \infty}{\sim} \text{Cste}\, q_0^{-3(2g-2+n)} dz_{1}\cdots dz_{n}.\eeq
Indeed, the recursion kernel behaves like  
\beq K(z_0,z)=\frac{(z^2-1)}{2z(z^2-z_0^2)(\theta-4q_0^3(z^2-1))}\frac{dz_{0}}{dz}=O\left(q_0^{-3}\right).\eeq
Thus adding $q_{0}^{-3}$ at each step of the recursion. In particular we find that:
\beq 
\underset{t\to \infty_{B,C}}{\lim} F^{(g)}_{\rm HTW}(t)=0 \quad \text{for $g \ge 2$}.
\eeq

\subsection{Tau-function and symplectic invariants for the HTW pair}
In this section we state the second main result of this paper. 

\begin{thm}\label{propo2} 
The HTW pair is of topological type (in the sense of Section \ref{DeterForm}) and we have:
\beq \label{eq:fg-and-tau-HTW-theorem}
\frac{dF_{\rm HTW}^{(g)}(t)}{dt}= 
- \sigma_{2g}(t) \quad \text{for $g \ge 2$}.
\eeq
\end{thm}

In particular we can verify from \eqref{eq:free-energy-HTW} that \eqref{eq:fg-and-tau-HTW-theorem} is correct for $g=2$ and $g=3$. Proof of Theorem \ref{propo2} will be given in Section \ref{DeterForm} and in appendix. We can also verify that 
\begin{equation}
\frac{dF_{\rm HTW}^{(0)}(t)}{dt} = -\left(\sigma_{0}(t) + \frac{t^2}{8}\right), \quad
\frac{dF_{\rm HTW}^{(1)}(t)}{dt} = -\sigma_{2}(t) 
\end{equation}
holds in accordance with \eqref{impor-HTW}. This leads to the following theorem: 

\begin{thm}\label{FgHTW}
The generating function of symplectic invariants of the HTW curve \eqref{SpecCurveHTW2} gives a $\tau$-function of Painlev\'e $2$. In other words, 
\beq
\ln \tau_{\text{\rm HTW}} = \sum_{g=0}^{\infty} \hbar^{2g-2} F_{\rm HTW}^{(g)}(t)  
\eeq
satisfies \eqref{impor-HTW}. Furthermore, in the both of limits
$t \to \infty_{B}$ and $t \to \infty_{C}$ we have:
\begin{equation} \label{eq:FgHTW-and-inftyB}
F_{\rm HTW}^{(g)}(t)=- \int_{\infty_{B,C}}^t \sigma_{2g}(s) ds 
\quad \text{for $g \ge 2$}.
\end{equation}
\end{thm}

\subsection{Limit at $t=\infty_A$: the Bessel curve}
\label{section:Limit-HTW-to-Bessel}

Let us realize the following affine symplectic transformation on the spectral curve \eqref{SpecCurveHTW}:
\beq x=\frac{2q_0^2}{\theta^2}X\,\,\,,\,\,\, y=\frac{\theta^2}{2q_0^2}Y \eeq
we get the new spectral curve:
\beq \label{eq:HTW-curve-before-limit} Y^2=\frac{\left(X+\theta^2\right)\left(1-\frac{4q_0^3}{\theta^3}X\right)^2}{4X^2}. \eeq
When $t\to \infty_{A}$ (i.e., $q_0\to 0$) we get that the limiting curve becomes 
\beq \label{Bessel} Y^2=\frac{X+\theta^2}{4X^2}, \eeq
what we call the Bessel curve. It can be parametrized into:
\beq \label{SpecCurveBessel} \left\{
\begin{array}{l}
 \displaystyle X(z)=\theta^2(z^2-1),  \\[+.3em]
 \displaystyle Y(z)=\frac{z}{2\theta(z^2-1)}. 
\end{array}
\right.\eeq

In particular straightforward computations of the topological recursion gives:
\bea \label{FgBessel}
F_{\text{Bessel}}^{(0)}&=&
-\frac{3\theta^2}{4} - \frac{\theta^2}{4}  
\ln \left(- \frac{1}{4\theta^2} \right)
\cr
F_{\text{Bessel}}^{(1)}&=&
- \frac{1}{24} \ln \left( - \frac{1}{4 \theta^2} \right), \cr
F_{\text{Bessel}}^{(2)}&=&-\frac{1}{240\,\theta^2},\nonumber \\[+.1em]
F_{\text{Bessel}}^{(3)}&=&\frac{1}{1008\,\theta^4}. 
\eea 
General properties regarding limits and symplectic transformations of the curve in the topological recursion tell us that:
\beq 
\underset{t\to \infty_{A}}{\lim} F_{\text{HTW}}^{(g)}(t)=F^{(g)}_{\text{Bessel}} \quad \text{for $g \ge 2$}. \eeq
On the other hand, it follows from \eqref{eq:int-sigma-Bernoulli} and \eqref{eq:FgHTW-and-inftyB} that we have
\begin{equation}
\underset{t\to \infty_{A}}{\lim} F_{\text{HTW}}^{(g)} =
- \int_{\infty_{B, C}}^{\infty_A} \sigma_{2g}(t) 
=\frac{B_{2g}}{2g(2g-2)\theta^{2g-2}} \quad \text{for $g \ge 2$}.
\end{equation}
Therefore, as a corollary of our main theorems, we have computed $F^{(g)}_{\rm Bessel}$ explicitly:
\begin{thm} \label{thm:Bessel-Fg} The symplectic invariants $F^{(g)}_{\rm Bessel}$ of the Bessel curve \eqref{SpecCurveBessel} are given by 
\beq \label{ByProduct} 
F^{(g)}_{\rm  Bessel} =\frac{B_{2g}}{2g(2g-2)\theta^{2g-2}} 
\quad \text{for $g \ge 2$}. \eeq
\end{thm}
To our knowledge, \eqref{ByProduct} has not been mentioned in the literature. 

\begin{rem} \label{remark:Yamada}
Y. Yamada pointed out to the authors that the JM curve 
and HTW curve are related by the following symplectic transformation:
\begin{eqnarray} \label{eq:Yamada-symplectic-transform}
x_{\rm HTW} = x_{\rm JM}^2 + y_{\rm JM} + \frac{t}{2}, \quad  
y_{\rm HTW} = - x_{\rm JM} + \frac{\theta}{2x_{\rm JM}^2 + 2y_{\rm JM} + t},
\end{eqnarray} 
where $(x_{\rm JM/HTW}, y_{\rm JM/HTW})$ are coordinates in the expressions \eqref{SpecCurveJM} and \eqref{SpecCurveHTW} of JM/HTW curve. Note that the parametrizations \eqref{SpecCurveJM2} and \eqref{SpecCurveHTW2} of JM/HTW curve are not compatible with this symplectic transformation. As is shown in \cite{EO-xy-symmetry}, a class of symplectic transformations (which includes the transformation $x \leftrightarrow y$) do not preserve $F^{(g)}$ in general; this explains why there is a discrepancy between $F^{(g)}_{\rm JM}$ and $F^{(g)}_{\rm HTW}$. Our computation shows that the difference is explicitly described by the Bernoulli numbers: 
\begin{equation} \label{eq:JM-HTW-Section-4}
F^{(g)}_{\rm JM} - F^{(g)}_{\rm HTW} = - \frac{B_{2g}}{2g(2g-2)\theta^{2g-2}} \quad \text{for $g \ge 2$}.
\end{equation}
In Appendix \ref{AppendixE} we show that the r.h.s. of \eqref{eq:JM-HTW-Section-4} coincides with the integration constant computed in \cite{EO-xy-symmetry}.
\end{rem}


\section{\label{DeterForm}Determinantal formulas and topological type property}

In this section we review the determinantal formulas formalism and the issue of the topological type property. Then we mention our main results and discuss about the consequences.  The proof of the topological type properties are postponed in Subsection \ref{AppendixC},  Appendix \ref{AppendixB} and \ref{AppendixA}.

\subsection{Determinantal formulas}

Here we remind the reader about determinantal formulas developed in \cite{Deter}. 

Determinantal formulas are built from a solution 
\begin{equation}
\Psi(x,t) =\begin{pmatrix} 
\psi(x,t) & \phi(x,t) \\ \td{\psi}(x,t) &\td{\phi}(x,t) \end{pmatrix}
\end{equation}
of the JM or the HTW Lax pair:
\beq \label{eq:differential-system-5} 
\hbar \frac{\partial}{\partial x} \Psi =\mathcal{D} \Psi \,\,,\,\, 
\hbar \frac{\partial}{\partial t} \Psi =\mathcal{R} \Psi 
\eeq
where the matrices $\mathcal{D}$ and $\mathcal{R}$ are traceless and have formal series expansions
\beq
{\mathcal D}(x,t) = \sum_{k=0}^{\infty} D^{(k)}(x,t) \hbar^{k} \,\,,\,
{\mathcal R}(x,t) = \sum_{k=0}^{\infty} R^{(k)}(x,t) \hbar^{k}
\eeq
We take a matrix-type WKB formal solution 
\beq \label{eq:WKB-solution}
\Psi(x,t) = \left( \sum_{k=0}^{\infty} \Psi^{(k)}(x,t)\hbar^{k} \right)\exp\left( \frac{T(x,t)}{\hbar} \right),\quad
T(x,t) = {\rm diag}(s(x,t), -s(x,t))
\eeq
which is normalized by $\det \Psi=1$. Here the phase function $s(x,t)$ satisfies
\beq \frac{\partial}{\partial x}s(x,t) = \sqrt{E_{\infty}(x,t)}, \eeq
where 
\bea \label{eq:E-infinity}
E_{\infty}(x,t) & = &  - \det {\mathcal D}^{(0)}(x,t) \nonumber \\[+.2em]
& = & \left\{
\begin{array}{ll}
\displaystyle (x-q_0)^2(x^2+2q_0x+3q_0^2+t) & \text{for JM case},\\[+.3em]
\displaystyle \frac{1}{2x^2}\left(x-\frac{\theta}{2q_0}\right)^2\left(x+2q_0^2\right) & \text{for HTW case}.
\end{array} \right. 
\eea

Determinantal formulas are obtained from the Christoffel-Darboux kernel:
\beq \label{PKK} 
K(x_1,x_2)=
\frac{\psi(x_1)\td{\phi}(x_2)-\td{\psi}(x_1)\phi(x_2)}{x_1-x_2}
\eeq
(here we are omitting the $t$-dependence for simplicity)
with the following definition:
\begin{defn}[Definition 2.3 of {\cite{Deter}}]
\label{DeterFormm} The (connected) correlation functions are defined by:
\bea \label{eq:W1}
W_1(x)&=&\frac{\partial \psi}{\partial x}(x)\td{\phi}(x)-
\frac{\partial \td{\psi}}{\partial x}(x)\phi(x),\cr
W_n(x_1,\dots,x_n)&=&-\frac{\delta_{n,2}}{(x_1-x_2)^2}+(-1)^{n+1}\sum_{\sigma: \text{$n$-cycles}}\prod_{i=1}^n K(x_i,x_{\sigma(i)}) \cr 
&   & \text{for $n \ge 2$}. \eea
\end{defn}

The correlation functions are formal power series of $\hbar$ whose coefficients are symmetric functions of $x_{1}, \dots, x_{n}$.
Note that there exists an alternative expression for the correlation functions \cite{Deter}. Define the rank $1$ projector by 
\beq \label{defM} M(x,t)=\Psi(x,t)\begin{pmatrix}1&0\\0&0\end{pmatrix} \Psi^{-1}(x,t)=\begin{pmatrix}\psi\td{\phi}&-\psi\phi\\ \td{\psi}\td{\phi}&-\phi\td{\psi}\end{pmatrix}.\eeq
It is in fact the canonical projector on the first coordinate taken into the basis defined by $\Psi(x,t)$. 
The rank $1$ projector satisfies:
\beq M^2=M \,\,,\,\, \Tr M=1 \,\,,\,\, \det M=0. \eeq
Theorem $2.1$ of \cite{Deter} gives an alternative expression for $W_n(x_1,\dots,x_n)$:
\bea \label{alternative} 
W_1(x)&=&-\frac{1}{\hbar} \Tr (\mathcal{D}(x)M(x)),\cr
W_2(x_1,x_2)&=&\frac{\Tr (M(x_1)M(x_2)) -1}{(x_1-x_2)^2},\cr
W_n(x_1,\dots,x_n)&=& (-1)^{n+1}\Tr \sum_{\sigma: \text{$n$-cycles}} \prod_{i=1}^n \frac{M(x_{\sigma(i)})}{x_{\sigma(i)}-x_{\sigma(i+1)}}\cr
&=&\frac{(-1)^{n+1}}{n}\sum_{\sigma \in S_n} \frac{ \Tr M(x_{\sigma(1)})\dots M(x_{\sigma(n)})}{(x_{\sigma(1)}-x_{\sigma(2)})\dots (x_{\sigma(n-1)}-x_{\sigma(n)})} \cr 
& &  \text{for $n \ge 3$}. 
\eea

\subsection{Topological type property}\label{subsection:TTP}

Now we give the definition of the topological type property for the differential equation \eqref{eq:differential-system-5} having the spectral curve of genus $0$.

\begin{defn}[Definition 3.3 of \cite{BBEnew}, 
Section 2.5 of \cite{Deter}] \label{def:TT-property-Lax}
The differential system \eqref{eq:differential-system-5} is said to be of topological type if the correlation functions $W_{n}$ given in Definition \ref{DeterFormm} satisfy the following three conditions: 
\begin{enumerate} 
\item[(1)] \underline{Parity property}: $W_{n}|_{\hbar \hspace{+.1em} \mapsto - \hbar} = (-1)^{n} W_{n}$ hold for $n \ge 1$. 

\smallskip
\item[(2)] \underline{Leading order property}: The leading order of the series expansion of the correlation function $W_n$ is at least of order $\hbar^{n-2}$. 
When these two conditions are satisfied, $W_{n}$ has the following expansion (called a topological expansion):
\beq \label{Exponents} W_n(x_{1},\dots,x_{n})=\sum_{g=0}^\infty \hbar^{2g-2+n}
W_n^{(g)}(x_1,\dots,x_n) \quad \text{for $n \ge 1$}. \eeq

\item[(3)] \underline{Pole property}: 
The coefficients $W_n^{(g)}(x_1,\dots,x_n)$ of correlation functions $W_n$ have no poles at even zeros of $E_\infty(x)$ given in \eqref{eq:E-infinity}.
\end{enumerate} 
\end{defn}

The authors of \cite{BBEnew} and \cite{Deter} proved that:
\begin{prop}[Theorem $2.1$ of \cite{Deter}, 
Theorem 3.1 and Corollary $4.2$ of \cite{BBEnew}]
\label{Conditions} ~
\begin{itemize}
\item[(i)]
If the differential system \eqref{eq:differential-system-5} is of topological type, then the coefficients $W_n^{(g)}(x_{1},\dots,x_{n})$ appearing in the expansion \eqref{Exponents} of the correlation function $W_{n}(x_{1},\dots,x_{n})$ are identical to Eynard-Orantin differentials $\omega^{(g)}_{n}(z_1,\dots,z_n)$ obtained from the topological recursion applied on the spectral curve $y^2=E_\infty(x)$ in the following way:
\begin{multline} \label{eq:EO-and-Wgn}
W^{(g)}_{n}(x(z_{1}),\dots,x(z_{n}))dx(z_{1})\cdots dx(z_{n}) = \omega^{(g)}_{n}(z_{1},\dots,z_{n}) \\ 
 \text{for $g \ge 0$ and $n \ge 1$},
\end{multline}
where $x(z)$ is the rational function of $z$ appearing in the parametrization \eqref{SpecCurveJM2} or \eqref{SpecCurveHTW2} of the spectral curve. 

\smallskip
\item[(ii)] 
Suppose that the differential system \eqref{eq:differential-system-5} is of topological type. Then, the generating function 
\[
\sum_{g = 0}^{\infty} \hbar^{2g-2} F_g(t)
\]
of the symplectic invariants $F_g$ obtained from the topological recursion applied to $y^2=E_\infty(x)$ gives the isomonodromic tau-function associated with \eqref{eq:differential-system-5} in the sense of Jimbo-Miwa-Ueno {\rm (\cite{JMI})}. 

\end{itemize}
\end{prop}

Here we omit the general definition of the Jimbo-Miwa-Ueno's isomonodromic tau-function (see Section 5 of \cite{JMI}; see also Section 4.2 in \cite{BBEnew} and Section 1.5 in \cite{P2WithoutCoeffFull}). In the case of JM/HTW Lax pair, the isomonodromic tau-function $\tau_{\rm JM/HTW}$ is defined (up to constant) by 
\begin{eqnarray} \label{eq:isomonodromic-tau-function} 
2 \Res_{x=\infty} \left(\frac{1}{\hbar} \frac{\partial s_{\infty}}{\partial t}(x) W_{1}(x) dx \right) = 
\left\{ 
\begin{array}{ll}
\displaystyle 
\frac{d}{dt} \ln \tau_{\rm JM}(t,\hbar) & \text{for JM case}, 
\\[+.9em] 
\displaystyle 
\frac{d}{dt} \ln \tau_{\rm HTW}(t,\hbar) & \text{for HTW case}.
\end{array} \right.
\end{eqnarray} 
where $W_{1}(x)$ is given in \eqref{eq:W1}, and $s_{\infty}(x)$ is the divergent part of $s(x)$ given in \eqref{eq:WKB-solution} when $x\rightarrow \infty$:
\begin{eqnarray} \label{eq:s-infty-JMHTW}
s_{\infty}(x) = \left\{
\begin{array}{ll}
\displaystyle 
\frac{x^{3}}{3} + \frac{t x}{2} & 
\text{for JM case}, \\[+.5em]
\displaystyle
\frac{\sqrt{2}x^{\frac{3}{2}}}{3} - 
\frac{t x^{\frac{1}{2}}}{\sqrt{2}} & 
\text{for HTW case}.
\end{array}
\right.
\end{eqnarray}
We can verify that the above definition of the tau-function is consistent with \eqref{impor} and \eqref{impor-HTW}, respectively, since $W_1(x)$ behaves as
\begin{eqnarray}
W_{1}(x) = \left\{
\begin{array}{ll}
\displaystyle \frac{x^{2}}{\hbar} + \frac{t}{2\hbar}- \frac{\theta}{\hbar x} + \frac{\sigma(t)}{\hbar x^{2}} + O(x^{-2}) & \text{for JM case}, \\[+.7em] 
\displaystyle \frac{x^{\frac{1}{2}}}{\sqrt{2} \hbar} - \frac{t x^{-\frac{1}{2}}}{2\sqrt{2}\hbar} - \frac{\sigma(t) + \frac{t^{2}}{8}}{\sqrt{2} \hbar}x^{-\frac{3}{2}} + O(x^{-2}) & \text{for HTW case}.
\end{array} \right.
\end{eqnarray}
when $x \rightarrow \infty$. As is explained in \cite{BBEnew, P2WithoutCoeffFull}, the claim (ii) in Proposition \ref{Conditions} is a consequence of the claim (i) and the ``variation formula" established in \cite{EO}. To make our paper self-contained, we will explain this point in Appendix \ref{AppendixD}.

Our context fits perfectly with Proposition \ref{Conditions} 
and thus only the proof of the topological type property remains. 
Our main theorem (including the statement of Theorem \ref{thm:JMFg-tau} 
and \ref{FgHTW}) claims that the conditions $(1) \sim (3)$ in 
Definition \ref{def:TT-property-Lax} hold for both JM and HTW Lax pairs:

\begin{thm}\label{MainTheo} For any choice of the monodromy parameter $\theta \ne 0$, both of the Jimbo-Miwa Lax pair \eqref{JMLaxPair} and the Harnad-Tracy-Widom Lax pair \eqref{HTWLaxPair} are of topological type. Therefore, the $\hbar$-expansion of the tau-function $\tau_{2g}$ and correlation functions $W_n^{(g)}$ respectively identify with the symplectic invariants $F^{(g)}$ and the Eynard-Orantin differentials $\omega_n^{(g)}$ computed from the topological recursion applied to the corresponding spectral curves \eqref{SpecCurveJM2} and \eqref{SpecCurveHTW2}. 
\end{thm}

We will prove that the three conditions $(1) \sim (3)$ in Definition \ref{def:TT-property-Lax} hold for both Lax pairs in the rest of Section \ref{DeterForm} and Appendix \ref{AppendixB}, \ref{AppendixA}.  We note that, in the previous works \cite{BBEnew,Deter,P5}, the leading order property (2) of correlation function $W_n$ was derived by using so-called ``insertion operator". However, we find that there is an incompleteness in the proof using the insertion operators. In next subsection we give another proof of the leading order property without using the insertion operator. Our new method shows that the leading order property (2) follows from the pole property (3). The parity property (1) and the no pole property (3) will be proved in Appendix \ref{AppendixB} and \ref{AppendixA}, respectively. We will also explain why the proof based on the insertion operator is incomplete in Appendix \ref{Incomplete}.

\subsection{\label{AppendixC}Proof of the leading order property of $W_n$ using loop equations}
This subsection is dedicated to prove that the both JM and HTW Lax pair enjoy the property (2) in Definition \ref{def:TT-property-Lax}; that is, the leading order of the series expansion in $\hbar$ of $W_n$ is at least of order $\hbar^{n-2}$: 
\beq \label{eq:order-n-2-Wn} W_n(x_{1},\dots,x_{n}) = O(\hbar^{n-2})\quad \text{for $n \ge 1$}. \eeq

Determinantal formulas in Definition \ref{DeterFormm} have been introduced so that they satisfy a set of equations known as the loop equations. These loop equations (also known as Schwinger-Dyson equations) originate in random matrix theory where they are crucial. We recall here the main result of \cite{Deter}:

\begin{prop}[Theorem $2.9$ of \cite{Deter}] Let us define the following functions (we denote by $L_n$ the set of variables $\{x_1,\dots,x_n\}$):
\bea 
P_1(x)&=&\frac{1}{\hbar^2}\det {\mathcal{D}}(x,t),\cr
P_2(x;x_2)&=&
\frac{1}{\hbar} 
{\Tr}\left( \frac{\mathcal{D}(x,t)-{\mathcal{D}}(x_2,t)-(x-x_2){\mathcal{D}}'(x_2,t)}{(x-x_2)^2}M(x_2)\right),\cr
P_{n+1}(x;L_n)&=&
(-1)^n\left[Q_{n+1}(x;L_n)-\sum_{j=1}^n 
\frac{1}{x-x_j}\Res_{x'\to x_j} Q_{n+1}(x',L_n)\right],\cr
Q_{n+1}(x;L_n) \cr 
& & \hspace{-6.em} = \frac{1}{\hbar}\sum_{\sigma\in S_n} 
\frac{{\Tr}\left( \mathcal{D}(x)M(x_{\sigma(1)})\dots 
M(x_{\sigma(n)})\right)}{(x-x_{\sigma(1)})
(x_{\sigma(1)}-x_{\sigma(2)})\dots 
(x_{\sigma(n-1)}-x_{\sigma(n)})(x_{\sigma(n)}-x)}. \cr & &
\eea
Then, the correlation functions satisfy 
\beq \label{eq:first-loop-equation} P_1(x)=W_2(x,x)+W_1(x)^2, \eeq
and 
\[
0=P_{n+1}(x;L_n)+W_{n+2}(x,x,L_n)+2W_1(x)W_{n+1}(x,L_n)
\]
\bea \label{ind1} 
+ \sum_{J \subset L_n, J\notin\{ \emptyset,L_n\}} W_{1+|J|}(x,J)W_{1+n-|J|}(x,L_n\setminus J) \nonumber \\
+ \sum_{j=1}^n \frac{d}{dx_j} \frac{ W_n(x,L_n\setminus x_j)-W_n(L_n)}{x-x_j} \quad \text{for $n \ge 1$}. 
\eea
Moreover $P_{n+1}(x;L_n)$ is a rational function of $x$ whose poles are at the poles of $\mathcal{D}(x,t)$.
\end{prop} 

The equations \eqref{eq:first-loop-equation} and \eqref{ind1} are called the loop equations. As we will see this proposition and a subtle induction are sufficient to prove that $W_n$ is at least of order $\hbar^{n-2}$. Let us now make the following crucial observation:

\begin{thm} In the JM Lax pair case, $P_{n+1}(x;L_n)$ does not depend on $x$ for $n\geq 1$. In the HTW Lax pair case, the functions $P_{n+1}(x;L_n)$ are of the form $P_{n+1}(x;L_n)= \frac{1}{x} \td{P}_{n+1}(L_n)$. 
\end{thm}

\begin{proof}
In the JM case, since the entries of $\mathcal{D}(x,t)$ is polynomial of $x$, $P_{n+1}(x;L_n)$ may only have singularity at $x=\infty$. However looking at large $x$ the definition shows that $P_{n+1}(x;L_n)$ can only be a polynomial of degree $0$ and hence does not depend on $x$. In addition we get an explicit formula:
\[
\hspace{-22.em}
P_{n+1}(x;L_n) = 
P_{n+1}(L_n) 
\]
\beq 
= \frac{(-1)^{n+1}}{\hbar}\sum_{\sigma\in S_n} \frac{\Tr\left( \sigma_3 M(x_{\sigma(1)})\dots M(x_{\sigma(n)})\right)}{(x_{\sigma(1)}-x_{\sigma(2)})\dots (x_{\sigma(n-1)}-x_{\sigma(n)})} \quad \text{for $n \ge 1$}. 
\eeq
For example, using directly the definition we have 
$P_2(x;x_2)=\frac{1}{\hbar}\Tr (\sigma_3 M(x_2))$ which is indeed independent of $x$. In the HTW case, the form of $\mathcal{D}(x,t)$ implies that $P_{n+1}(x;L_n)$ may only have simple poles at $x=0$ and a simple zero at infinity (degree of numerator-denominator shows that it behaves as 
$O\left({x}^{-1}\right)$ at infinity). Hence we conclude that it is proportional to $x^{-1}$ (one could even get a complete expression by taking the residue at $x=0$).
\end{proof}

We now have all the ingredients to prove the following theorem:

\begin{thm} \label{thm:main-appendix-C}
The correlation functions $W_n(x_1,\dots,x_n)$ admit a series expansion in $\hbar$ starting at least at order $\hbar^{n-2}$.
\end{thm}

The rest of Section \ref{AppendixC} is devoted to the proof of Theorem \ref{thm:main-appendix-C}. Our proof is done by induction. We will denote $L_i=\{x_1,\dots,x_i\}$. Let us define the following statement:
\beq \mathcal{P}_k \,:\,  
W_j(x_1,\dots,x_j) \text{ is at least of order } \hbar^{k-2} \quad \text{for $j \ge k$}. \eeq

We can easily verify that the correlation functions $W_n$ for $n \ge 2$ has the formal series expansion of the form 
\begin{equation} \label{eq:small-w-n-k}
W_n(x_1,\dots,x_n) = \sum_{k=0}^{\infty}w_n^{(k)}(x_1,\dots,x_n)
\hbar^{k} \text{ for }n\geq 2.
\end{equation}
It also follows from the first loop equation \eqref{eq:first-loop-equation} that $W_1(x)$ has the formal series expansion of the form 
\begin{equation} \label{eq:small-w-1-k}
W_1(x)=\sum_{k=-1}^\infty w_1^{(k)}(x)\hbar^k, 
\end{equation}
where the leading term $w_1^{(-1)}(x)$ coincides with $\sqrt{E_{\infty}(x)}$. Indeed, let us look the first loop equation \eqref{eq:first-loop-equation}. Since $W_2(x_1,x_2)$ only starts at $\hbar^0$ while $P_1(x)= \frac{1}{\hbar^2}\det \mathcal{D}$ looking at order $\hbar^{-2}$ in the last equation provides the result. Thus we have checked that the statement ${\mathcal P}_1$ is true. The statement ${\mathcal P}_2$ is obviously true in view of \eqref{eq:small-w-n-k}.

Let us assume that the statement $\mathcal{P}_i$ is true for all $i\leq n$ for some integer $n \ge 2$. Now we look at the loop equation \eqref{ind1}. By the induction hypothesis, we have that the last two terms are at least of order $\hbar^{n-2}$. Indeed in the sum we have terms of order $\hbar^{1+|J|-2+1+n-|J|-2}=\hbar^{n-2}$. Moreover we also have from the same assumption that $W_{n+2}(x,x,L_n)$ is also of order at least $\hbar^{n-2}$ (since $n+2\geq n$). Eventually $W_{n+1}(x,L_n)$ is at least of order $\hbar^{n-2}$ so we get when looking at order $\hbar^{n-3}$ in \eqref{ind1}:
\beq \label{eqsys} 0=P_{n+1}^{(n-3)}(x;L_n)+2w_1^{(-1)}(x)w_{n+1}^{(n-2)}(x,L_n). \eeq
Here $P_{n+1}^{(\ell)}(x;L_n)$ is the coefficient of $\hbar^{\ell}$ in the $\hbar$-expansion of $P_{n+1}(x;L_n)$. If we assume that $w_{n+1}^{(n-2)}(x,L_{n})\neq 0$ then we have:
\beq w_{n+1}^{(n-2)}(x,L_{n})=\frac{P_{n+1}^{(n-3)}(x;L_{n})}{2w_1^{(-1)}(x)}. \eeq
In our cases we get:
\begin{eqnarray}
w_{n+1}^{(n-2)}(x,L_{n}) = \left\{ 
\begin{array}{ll}
\frac{P_{n+1}^{(n-3)}(L_{n})}{2(x-q_0)\sqrt{(x+q_0)^2+\frac{\theta}{q_0}} } \quad \text{for JM case}, \\[+1.em]
\frac{\td{P}_{n+1}^{(n-3)}(L_{n})}{\left(x-\frac{\theta}{2q_0}\right)\sqrt{x+2q_0^2}} \quad \hs{+2.5em} \text{for HTW case}. 
\end{array} \right.
\end{eqnarray}
In both cases we obtain that $w_{n+1}^{(n-2)}(x,L_{n})$ must have a simple pole at the even zero of $E_{\infty}(x)$, and this contradicts the pole property which will be  proved in Appendix \ref{AppendixA}. Consequently we must have $w_{n+1}^{(n-2)}(x,L_{i_0})=0$. This proves that $w_{n+1}(x,L_n)$ is at least of order $\hbar^{n-1}$. 

To complete the proof of ${\mathcal P}_{n+1}$, we now need to prove the same statement for higher correlation functions. Let us prove it by a second induction by defining:
\beq \td{\mathcal{P}}_i \,:\, W_i(x_1,\dots,x_i) \text{ is of order at least } \hbar^{n-1}. \eeq
We want to prove $\td{\mathcal{P}}_i$ for all $i\geq n+1$ by induction. We just proved it for $i=n+1$ so initialization is done. Let us assume that $\td{\mathcal{P}}_j$ is true for all $j$ satisfying $n+1\leq j\leq i_0$. We look at the loop equation:
\bea \label{ind2} 0&=&P_{i_0+1}(x;L_{i_0})+W_{i_0+2}(x,x,L_{i_0})+2W_1(x)W_{i_0+1}(x,L_{i_0})\cr
&&+ \sum_{J \subset L_{i_0}, J\notin\{ \emptyset,L_{i_0}\}} W_{1+|J|}(x,J)W_{1+i_0-|J|}(x,L_{i_0}\setminus J)\cr
&&+\sum_{j=1}^{i_0} \frac{d}{dx_j} \frac{ W_{i_0}(x,L_{i_0}\setminus x_j)-W_{i_{0}}(L_{i_0})}{x-x_j}. 
\eea
By assumption on $\td{\mathcal{P}}_{i_0}$, the last sum with the derivatives contains terms of order at least $\hbar^{n-1}$. In the sum involving the subsets of $L_{i_0}$ it is straightforward to see that the terms are all of order at least $\hbar^{n-1}$. Indeed, as soon as one of the index is greater then $n+1$, the assumption of $\td{\mathcal{P}}_i$ for $n+1\leq i\leq i_0$ tells us that this term is already at order at least $\hbar^{n-1}$. Since the second factor of the product is at least of order $\hbar^0$ then it does not decrease the order. Now if both factors have indexes strictly lower than $n+1$, then the assumption of $\mathcal{P}_j$  for all $j\leq n$ tell us that the order of the product is at least of $\hbar^{|J|+1-2+1+i_0-|J|-2}=\hbar^{i_0-2}$ which is greater than $n-1$ since $i_0\geq n+1$. Additionally by induction on $\mathcal{P}_n$ we know that $W_{i_0+1}(x,L_{i_0})$ is at least of order $\hbar^{n-2}$ as well as $W_{i_0+2}(x,x,L_{i_0})$. Consequently looking at order $\hbar^{n-3}$ in \eqref{ind2} gives:
\beq \label{eqsys2} 0=P_{i_0+1}^{(n-3)}(x;L_{i_0})+2w_1^{(-1)}(x)w_{i_0+1}^{(n-2)}(x,L_{i_0}). \eeq
We can apply a similar reasoning as for \eqref{eqsys}. If we assume that $w_{i_0+1}^{(n-2)}(x,L_{i_0})\neq 0$, then we have:
\beq w_{i_0+1}^{(n-2)}(x,L_{i_0})=\frac{P_{i_0+1}^{(n-3)}(x;L_{i_0})}{2w_1^{(-1)}(x)}. \eeq
In our two cases we get:
\begin{eqnarray}
w_{i_0+1}^{(n-2)}(x,L_{i_0}) = \left\{ 
\begin{array}{ll}
\frac{P_{i_0+1}^{(n-3)}(L_{i_0})}{2(x-q_0)\sqrt{(x+q_0)^2+\frac{\theta}{q_0}} }  \quad \text{for JM case} \\[+1.em]
\frac{\td{P}_{i_0+1}^{(n-3)}(L_{i_0})}{\left(x-\frac{\theta}{2q_0}\right)\sqrt{x+2q_0^2} } \quad \hspace{+2.3em} \text{for HTW case}.
\end{array}
\right.
\end{eqnarray}
In both cases we obtain that $w_{i_0+1}^{(n-2)}(x,L_{i_0})$ must have a simple pole at the even zero of $E_{\infty}(x)$ which contradicts the pole property that will be proved in Appendix \ref{AppendixA}. Consequently we must have $w_{i_0+1}^{(n-2)}(x,L_{i_0})=0$. In particular it means that $w_{i_0+1}(x,L_{i_0})$ (which by assumption of $\mathcal{P}_n$ was already known to be of order $\hbar^{n-2}$) is at least of order $\hbar^{n-1}$ thus making the induction on $\td{\mathcal{P}}_{i_0}$. Hence by induction we have proved that $\forall\, i \geq n+1$, $\td{\mathcal{P}}_{i}$ holds which exactly proves that $\mathcal{P}_{n+1}$ is true. Eventually by induction we have just proved that $\mathcal{P}_{n}$ holds for $n \ge 1$, which implies the desired property \eqref{eq:order-n-2-Wn}. 

\begin{rem}
Several important observations can be made about this proof:
\begin{itemize}
\item The proof heavily relies on the pole property of the correlation functions $W_n^{(g)}$. In particular it is central to know that the correlation functions are regular at the even zeros of $E_{\infty}(x)$ since it provides the contradiction in \eqref{eqsys} and \eqref{eqsys2}.
\item The possible poles of $\mathcal{D}(x,t)$ are irrelevant in the proof. Indeed, they specify the form of $P_{n+1}(x;L_n)$ but do not play an important role in the contradiction of \eqref{eqsys} and \eqref{eqsys2}.
\item The presence of at least one even zero in the spectral curve is necessary in our proof because $\mathcal{D}(x,t)$ is a polynomial of degree $2$ in the JM case or has a simple pole at $x=0$ in the HTW case. In the case when $\mathcal{D}(x,t)$ is a polynomial of degree $1$, then $P_{n+1}(x;L_n)$ would be identically zero and thus our proof would also work in a simpler way. Hence the central element is the balance between the order of the singularity of $\mathcal{D}(x,t)$ and the fact that the spectral curve is of genus $0$.
\item This proof can be applied to more general cases: as soon as the spectral curve has a double zero and the pole structure is proved then the method can be applied. In the case of Lax pairs, the pole property is usually easy to obtain from the $t$-differential equation like we did in Appendix \ref{AppendixA} and the spectral curve is even simpler to obtain. 
\end{itemize}
\end{rem}

\section{Outlooks}

We believe that the topological type property (TT property) should hold more general Lax pairs including those for all six Painlev\'{e} equations with arbitrary (generic) monodromy parameters. The previous works in this direction treats Painlev\'e equation with a specific monodromy parameters (Cf. \cite{P2WithoutCoeffFull, P5}), but our work shows that the absence of monodromy parameter is not a necessary condition to obtain the TT property. Moreover, in this paper we have introduced new general methods to prove the TT property (see Section \ref{AppendixC} and Appendix \ref{AppendixA}) that should generalize easily for other Lax pairs. In fact, while this article was under peer review, the authors together with A. Saenz succeeded in proving the TT property for all six Painlev\'e equations (\cite{IMS}). Several natural questions arise from this work:
\begin{itemize} 
\item At the conceptual level a better understanding would definitely be an interesting development. Indeed it is even unclear so far to what extent the TT property is connected with Lax pairs. Does it work for any Lax pairs? Only specific integrable systems ? Our new methods seem general enough to work for many Lax pairs but a better understanding of the scope and limits of the methods is required.
\item On a totally different perspective, we have shown here that studying Lax pairs and their symplectic invariants may lead to explicit formulas for symplectic invariants of new spectral curves. Cases where general formulas for the symplectic invariants are known explicitly are extremely rare and this could provide a way to improve the classification of symplectic invariants for simple spectral curves. This knowledge might be of some use for enumerative geometry where a list of spectral curves and associated symplectic invariants would be helpful.
\item In this paper we also drew attention on the incompleteness of the insertion operator method. This calls for a better definition of the insertion operator to fix the current problem.
\end{itemize}


\appendix

\section{\label{AppendixB}Proof of the parity symmetry}

We want to prove that the $\hbar$ series expansion of the determinantal formulas $W_n(x_1,\dots,x_n)$ only involves powers of $\hbar$ of the same parity. (Cf., \eqref{Exponents}). In order to do this, we use Proposition $3.3$ of \cite{BBEnew} that gives a sufficient criteria to obtain the $\hbar\leftrightarrow -\hbar$ symmetry. We recall their proposition here:

\begin{prop}[Proposition $3.3$ of \cite{BBEnew}] Let us denote $\dagger$ the operator that change $\hbar$ into $-\hbar$. If there exists an invertible matrix $\Gamma(t)$ independent of $x$ such that:
\beq\label{eq:L-and-Gamma} 
\Gamma(t) \mathcal{D}^t(x,t)\Gamma^{-1}(t)=\mathcal{D}^\dagger(x,t),\eeq
then the correlators $W_{n}$ satisfy 
\beq W_{n}^{\dagger} = (-1)^{n} W_{n} \quad \text{for $n \ge 1$}. \eeq 
\end{prop}

In particular if this proposition is satisfied then it automatically follows that the $\hbar$ expansion a given function $W_n(x_1,\dots,x_n)$ may only involve powers of $\hbar$ with the same parity (given by the parity of $n$). Therefore all we have to do is prove the existence of a suitable matrix $\Gamma(t)$ for our two Lax pairs.

\medskip

Recall that $\sigma^{\dagger} = \sigma$ and $p^{\dagger} = p$ hold (see \eqref{tauexp} and  \eqref{pexp}). Then, it follows from \eqref{DeformedEq} that $q$ satisfies 
\begin{equation} \label{eq:q-dagger}
q^{\dagger} = - q - \frac{\theta}{p}. 
\end{equation}
Using this relation we can obtain $\mathcal{D}^\dagger(x,t)$ and we can find an invertible matrix $\Gamma$ satisfying \eqref{eq:L-and-Gamma} as follows:

\begin{thm} \label{prop:parity-property}We can find suitable matrices $\Gamma(t)$ for our Lax pairs:
\begin{itemize}
\item For the Jimbo-Miwa case, the matrix 
\beq
\Gamma(t) = \begin{pmatrix}1&0\\0&-2p(t)\end{pmatrix}
\eeq
satisfies \eqref{eq:L-and-Gamma}. 
\item For the Harnad-Tracy-Widom case, the matrix 
\beq
\Gamma(t) = \begin{pmatrix}1& \frac{p(t)}{2\theta} \\ 
\frac{p(t)}{2\theta}&0\end{pmatrix}
\eeq 
satisfies \eqref{eq:L-and-Gamma}. 
\end{itemize}
Consequently, the series expansion in $\hbar$ for $W_n$ only involves even (resp. odd) powers of $\hbar$ when $n$ is even (resp. odd). 
\end{thm}
Theorem \ref{prop:parity-property} follows from \eqref{eq:q-dagger} immediately. In Jimbo-Miwa case we have
\[
\mathcal{D}^{\dagger}(x,t) = 
\begin{pmatrix} x^2 + p + \frac{t}{2} & x+q+\frac{\theta}{p} \\ 
-2p(x-q) & -\bigl(x^2 + p + \frac{t}{2}\bigr) \end{pmatrix} 
\]
and 
\[
\mathcal{D}^t(x,t) = \begin{pmatrix} x^2 + p + \frac{t}{2} &-2p\bigl(x+q+\frac{\theta}{p}\bigr)  \\[+.4em] 
x-q & -\left(x^2 + p + \frac{t}{2}\right) \end{pmatrix}  ,
\]
while in the Harnad-Tracy-Widom case we have:
\[ 
\mathcal{D}^{\dagger}(x,t) = 
\begin{pmatrix} q+\frac{\theta}{p}+\frac{\theta}{2x} & x-p-2\bigl(q+\frac{\theta}{p}\bigr)^2-t \\ 
\frac{1}{2}+\frac{p}{2x} & -\bigl(q+\frac{\theta}{p}+\frac{\theta}{2x}\bigr) \end{pmatrix} 
\]
and 
\[
\mathcal{D}^t(x,t) = \begin{pmatrix} -q+\frac{\theta}{2x}&\frac{1}{2}+\frac{p}{2x}  \\ 
x-p-2q^2-t & q-\frac{\theta}{2x} \end{pmatrix}  .
\]
Then \eqref{eq:L-and-Gamma} can be checked easily by matrix multiplication in both cases.


\section{\label{AppendixA} Proof of the pole property}

As mentioned earlier we are interested in this article about solutions $q(t)$ of Painlev\'{e} $2$ admitting a formal expansion in the $\hbar$ parameter. Consequently for both Lax pairs, this implies a series expansion in $\hbar$ for $M(x,t)$, $W_n$, $\mathcal{D}(x,t)$ and $\mathcal{R}(x,t)$ of the form:
\bea \mathcal{D}(x,t)&=&\sum_{k=0}^\infty \mathcal{D}^{(k)}(x,t)\hbar^k \cr
\mathcal{R}(x,t)&=&\sum_{k=0}^\infty 
\mathcal{R}^{(k)}(x,t)\hbar^k,\cr
M(x,t)&=&\sum_{k=0}^\infty M^{(k)}(x,t)\hbar^k.
\eea
In this appendix an exponent $\,^{(k)}$ denotes the coefficient of $\hbar^{k}$ in the $\hbar$-expansion. Moreover, it is also obvious from the definitions of both Lax pairs that $\mathcal{D}^{(k)}(x,t)$ and $\mathcal{R}^{(k)}(x,t)$ do not depend on $x$ for $k\geq 1$.

\subsection{Jimbo-Miwa Lax pair}
We want to prove that the matrices $M^{(k)}(x,t)$ only have singularities (as a function of $x$) at the branch points of the JM spectral curve and possibly a pole at infinity. The plan is first to compute explicitly $M^{(0)}(x,t)$ and then find a recursive relation between the matrices.
\subsubsection{Computation of $M^{(0)}(x,t)$}
Inserting the series expansion of $M(x,t)$ into the differential system for $M(x,t)$:
\bea  \hbar\partial_x M(x,t)=\left[\mathcal{D}(x,t),M(x,t)\right] \\
\label{eq:differential-system-for-projector}
\hbar\partial_t M(x,t)=\left[\mathcal{R}(x,t),M(x,t)\right]
\eea
gives that:
\beq \label{Order0M}0=\left[\mathcal{D}^{(0)}(x,t),M^{(0)}(x,t)\right] \text{ and } 0=\left[\mathcal{R}^{(0)}(x,t),M^{(0)}(x,t)\right]. \eeq
Additionally, since $M(x,t)$ is a rank $1$ projector we know that $\Tr M(x,t)=1$ and $\det M(x,t)=0$. At order $\hbar^0$ this is equivalent to $\Tr M^{(0)}(x,t)=1$ and $\det M^{(0)}(x,t)=0$. Since $M^{(0)}(x,t)_{2,2}=1-M^{(0)}(x,t)_{1,1}$, the second equation of \eqref{Order0M} only gives $3$ different equations:
\bea 0&=&\frac{1}{2}\left(M^{(0)}(x,t)\right)_{2,1}-\frac{\theta}{2q_0}\left(M^{(0)}(x,t)\right)_{1,2},\cr
0&=&\frac{1}{2}\left(1-2\left(M^{(0)}(x,t)\right)_{1,1}\right)+(x+q_0)\left(M^{(0)}(x,t)\right)_{1,2},\cr
0&=&(x+q_0)\left(M^{(0)}(x,t)\right)_{2,1}+\frac{\theta}{2q_0}\left(1-2\left(M^{(0)}(x,t)\right)_{1,1}\right). 
\eea
We have used here the fact that $p_0=-\frac{\theta}{2q_0}$ and $t=-2q_0^2+\frac{\theta}{q_0}$. It is easy to observe that only two of the previous equations are independent. Therefore we have so far a system of $2$ independent equations with $3$ unknowns. In order to complete it, we must use the fact that $\det M^{(0)}(x,t)=0$. In the end we find the following system:
\bea 0&=&\frac{1}{2}\left(M^{(0)}(x,t)\right)_{2,1}-\frac{\theta}{2q_0}\left(M^{(0)}(x,t)\right)_{1,2},\cr
0&=&\frac{1}{2}\left(1-2\left(M^{(0)}(x,t)\right)_{1,1}\right)+(x+q_0)\left(M^{(0)}(x,t)\right)_{1,2},\cr
0&=&\left(M^{(0)}(x,t)\right)_{1,1}\left(1-\left(M^{(0)}(x,t)\right)_{1,1}\right) \cr & &-\left(M^{(0)}(x,t)\right)_{1,2}\left(M^{(0)}(x,t)\right)_{2,1}. 
\eea
It is also important to note that the first equation of \eqref{Order0M} would have lead exactly to the same system of equations. This system can be solved explicitly and we find:
\beq \label{M0JM} M^{(0)}(x,t)=\begin{pmatrix}\frac{1}{2}+\frac{x+q_0}{2\sqrt{\left(x+q_0\right)^2+\frac{\theta}{q_0}}}& \frac{1}{2\sqrt{\left(x+q_0\right)^2+\frac{\theta}{q_0}}}\\
\frac{\theta}{2q_0\sqrt{\left(x+q_0\right)^2+\frac{\theta}{q_0}}}&\frac{1}{2}-\frac{x+q_0}{2\sqrt{\left(x+q_0\right)^2+\frac{\theta}{q_0}}}\end{pmatrix}.
\eeq
It is obvious that $M^{(0)}(x,t)$ only have singularities at the branch points of the spectral curve. In particular, it is holomorphic at the even zero $x = q_0$ of $E_{\infty}(x)$ given by \eqref{eq:E-infinity}.

\subsubsection{Recursive system for higher orders}
\label{subsection:pole-str}

Since $\Tr M^{(k)}(x,t) = 0$ for $k \ge 1$, it suffices to 
consider the pole structure of 
$M^{(k)}(x,t)_{1,1}$, $M^{(k)}(x,t)_{1,2}$ and 
$M^{(k)}(x,t)_{2,1}$. 
Looking at order $\hbar^k$ with $k\geq 1$ in 
\eqref{eq:differential-system-for-projector},  
we have
\begin{eqnarray*}
\left[ \mathcal{R}^{(0)}(x,t), M^{(k)}(x,t)\right]=
\partial_t M^{(k-1)}(x,t) \hspace{+10.em} \\ 
-\underset{i=0}{\overset{k-1}{\sum}} 
\left[\mathcal{R}^{(k-i)}(x,t),M^{(i)}(x,t)\right] 
- \left[ \mathcal{R}^{(k)}(x,t), M^{(0)}(x,t) \right].
\end{eqnarray*}
Thus we get the following linear system
\[
\begin{pmatrix} 0&-\frac{\theta}{2q_0}&\frac{1}{2}\\
-1&x+q_0&0\\
x+q_0&\frac{\theta}{2q_0}&\frac{1}{2}\end{pmatrix}
\begin{pmatrix} M^{(k)}(x,t)_{1,1}\\ 
M^{(k)}(x,t)_{1,2}\\ M^{(k)}(x,t)_{2,1}\end{pmatrix} = 
\begin{pmatrix} J^{(k)}(x,t)_{1}\\ 
J^{(k)}(x,t)_{2}\\ J^{(k)}(x,t)_{3}\end{pmatrix}, 
\]
where $J^{(k)}(x,t)_{\ell}$ are polynomials which are
written in terms of $M^{(i)}(x,t)$ for $0 \le i \le k-1$ 
and their $t$-derivatives, and ${\RR}^{(i)}(x,t)$ $0 \le i \le k$. Since
\beq 
\det \begin{pmatrix} 0&-\frac{\theta}{2q_0}&\frac{1}{2}\\
-1&x+q_0&0\\
x+q_0&\frac{\theta}{2q_0}&\frac{1}{2}\end{pmatrix}=
-\frac{1}{2}\left((x+q_0)^2+\frac{\theta}{q_0}\right),
\eeq 
we get 
\[
\hspace{-13.em}
\begin{pmatrix}  
M^{(k)}(x,t)_{1,1}\\ M^{(k)}(x,t)_{1,2}\\ 
M^{(k)}(x,t)_{2,1}\end{pmatrix}=
\frac{1}{(x+q_0)^2+\frac{\theta}{q_0}} 
\]
\beq  \label{PolePropJM} \times 
\begin{pmatrix}
-(x+q_0)&-\frac{\theta}{q_0}& x+q_0\\
-1& x+q_0&1\\
2(x+q_0)^2+\frac{\theta}{q_0}& \frac{(x+q_0)\theta}{q_0}&
\frac{\theta}{q_0}
\end{pmatrix} 
\begin{pmatrix} J^{(k)}(x,t)_{1}\\ 
J^{(k)}(x,t)_{2}\\ J^{(k)}(x,t)_{3}\end{pmatrix}.
\eeq
Then, since we know that $\mathcal{R}^{(i)}(x,t)$'s are holomorphic at $x = q_0$ for all $i \ge 0$, a straightforward induction using \eqref{PolePropJM} shows that the only singularities of $M^{(k)}(x,t)$ are at the branch points and possibly a pole at infinity:

\begin{thm} \label{JMPoleStructure} For $k\geq 0$, the matrices $M^{(k)}(x,t)$ only have singularities at the branch points of the spectral curve $x=-q_0\pm \sqrt{{\theta}/{q_0}}$ and a possible pole singularity at infinity. In particular they are holomorphic at the even zero $x=q_0$ of $E_{\infty}(x)$ given by \eqref{eq:E-infinity}. Consequently the same singularity structure holds for the functions $W_n^{(g)}(x_1,\dots,x_n)$ thanks to the relation \eqref{alternative}.
\end{thm}

\begin{rem} \label{rem:pole-structure-and-t-derivative}
It is also interesting to observe that using the differential equation in $x$ rather than the one in $t$ provides a similar linear equation instead of \eqref{PolePropJM}:
\[
\hspace{-13.em}
\begin{pmatrix}  
M^{(k)}(x,t)_{1,1}\\ M^{(k)}(x,t)_{1,2}\\ 
M^{(k)}(x,t)_{2,1}\end{pmatrix}=
\frac{1}{(x-q_0)\left((x+q_0)^2+\frac{\theta}{q_0}\right)} 
\]
\[
\times 
\begin{pmatrix}
-(x+q_0)&-\frac{\theta}{q_0}& x+q_0\\
-1& x+q_0&1\\
2(x+q_0)^2+\frac{\theta}{q_0}& \frac{(x+q_0)\theta}{q_0}&
\frac{\theta}{q_0}
\end{pmatrix} 
\begin{pmatrix} \td{J}^{(k)}(x,t)_{1}\\ 
\td{J}^{(k)}(x,t)_{2}\\ \td{J}^{(k)}(x,t)_{3}\end{pmatrix}.
\]
However in this case, it is not easy to exclude the pole at the double zero $x=q_0$ of the spectral curve. Thus we understand here the importance of the differential equation with respect to $t$ in the context of the determinantal formulas. 
\end{rem}

\subsection{Harnad-Tracy-Widom Lax pair}
Most of the arguments of the previous section also apply to the Harnad-Tracy-Widom Lax pair.
\subsubsection{Computation of $M^{(0)}(x,t)$}
Looking at 
$\left[\mathcal{R}^{(0)}(x,t),M^{(0)}(x,t)\right]=0$ and $\det M^{(0)}(x,t)=0$ leads to the following system:
\bea 0&=&-x M^{(0)}(x,t)_{2,1}+\frac{1}{2}M^{(0)}(x,t)_{1,2},\cr
0&=&-x\left(1-2M^{(0)}(x,t)_{1,1}\right)+2q_0M^{(0)}(x,t)_{1,2},\cr
0&=&-2q_0M^{(0)}(x,t)_{2,1}+\frac{1}{2}\left(1-2M^{(0)}(x,t)_{1,1}\right),\cr
0&=&M^{(0)}(x,t)_{1,1}\left(1-M^{(0)}(x,t)_{1,1}\right)-M^{(0)}(x,t)_{1,2}M^{(0)}(x,t)_{2,1}. \cr & &
\eea
Note that only two of the first three equations are independent. This system of equations admits a unique solution given by:
\beq M^{(0)}(x,t)=\begin{pmatrix} \frac{1}{2}-\frac{q_0}{\sqrt{2}\sqrt{x+2q_0^2}}&\frac{x}{\sqrt{2}\sqrt{x+2q_0^2}}\\
\frac{1}{2\sqrt{2}\sqrt{x+2q_0^2}}&\frac{1}{2}+ \frac{q_0}{\sqrt{2}\sqrt{x+2q_0^2}}\end{pmatrix}.
\eeq
It is then obvious that $M^{(0)}(x,t)$ only have singularities at $x=-2q_0^2$ the unique branch point of the spectral curve and at infinity. Note also that $\left[\mathcal{D}^{(0)}(x,t),M^{(0)}(x,t)\right]=0$ would have provided an equivalent system of equations.

\subsubsection{Recursive system for higher orders}

By the same method presented in Subsection \ref{subsection:pole-str}, we can derive a linear equation satisfied by the entries of $M^{(k)}(x,t)$: 
\[
\begin{pmatrix} 0&\frac{1}{2}&-x\\
2x&2q_0&0\\
2q_0&-\frac{1}{2}&-x\end{pmatrix}
\begin{pmatrix} M^{(k)}(x,t)_{1,1}\\ M^{(k)}(x,t)_{1,2}\\ 
M^{(k)}(x,t)_{2,1}\end{pmatrix} = 
\begin{pmatrix} J^{(k)}(x,t)_{1}\\ 
J^{(k)}(x,t)_{2}\\ J^{(k)}(x,t)_{3}\end{pmatrix}, 
\]
where, as well as in the JM case,  
$J^{(k)}(x,t)_{\ell}$ are polynomials written in terms of 
$M^{(i)}(x,t)$ for $0 \le i \le k-1$ 
and ${\RR}^{(i)}(x,t)$ $0 \le i \le k$.
A straightforward computation shows that:
\beq \det \begin{pmatrix} 0&\frac{1}{2}&-x\\
2x&2q_0&0\\
2q_0&-\frac{1}{2}&-x\end{pmatrix}=2x(x+2q_0^2) \eeq
which is non-vanishing at $x={\theta}/({2q_0})$. 
Therefore, the same discussion given in 
Subsection \ref{subsection:pole-str} shows the following.

\begin{thm}\label{HTWPoleStructure} For $k\geq 0$, the matrices $M^{(k)}(x,t)$ only have singularities at the branch point of the spectral curve $x=-2q_0^2$ and poles at $x=0$ and $x=\infty$. In particular they are holomorphic at the even zero $x={\theta}/({2q_0})$ of $E_{\infty}(x)$ given by \eqref{eq:E-infinity}. Consequently the same singularity structure holds for the functions $W_n^{(g)}(x_1,\dots,x_n)$ thanks to the relation \eqref{alternative}. \end{thm}

Here we again note that the differential equation in $x$ 
is not helpful to show Theorem \ref{HTWPoleStructure} 
as well as in the JM case 
(see Remark \ref{rem:pole-structure-and-t-derivative}).


\section{\label{Incomplete}Incomplete proof using an insertion operator}
In \cite{BBEnew,Deter,P5} the various authors presented the construction of an insertion operator $\delta_\eta$ to prove the leading order of the $\hbar$ expansion of $W_n$. Unfortunately this proof is incomplete and requires an important modification to be correct that is currently being investigated. We present here the main reason for the incompleteness of this method. 

The method of the insertion operators naturally applies to the Picard-Vessiot (PV) ring $\mathbb{B}_1$ attached to the Lax pair, that is to say to the differential ring generated by the entries of $\Psi(x)$ and the scalar function $(\det \Psi(x))^{-1}$ over the differential ring ${\mathbb K}_1$ of rational functions of $x$. Taking an arbitrary large number of spatial variables, we end up with the projective limit: $\mathbb{B}_\infty=\bigcup_{i=1}^\infty \mathbb{B}_i$ over the field ${\mathbb K}_\infty=\bigcup_{i=1}^\infty {\mathbb K}_i$ of rational functions in any arbitrary large number of variables $x_i$. Most quantities defined in this paper belongs to the PV ring $\mathbb{B}_\infty$ since they can be expressed with the entries of $\Psi(x)$ and $(\det \Psi(x))^{-1}$. For example, matrix elements of $\Psi^{-1}(x)$, $\mathcal{D}(x)$, $\mathcal{R}(x)$, $K(x_1,x_2)$, $M(x)$ and all $W_n(x_1,\dots,x_n)$ belong to $\mathbb{B}_\infty$. The idea of the insertion operator is to create an operator $\delta_\eta$ acting on the PV ring that satisfies the following properties (i) $\sim$ (v)  (Cf. Definition 2.5, Definition 4.2 and Section 5.7.2 of \cite{BBEnew}):
\begin{itemize} 
\item[(i)] $\delta_{\eta} ({\mathbb K}_{\infty}) = 0$, and  $\delta_{\eta} ({\mathbb B}_{n}) \subset {\mathbb B}_{n+1}$.
\item[(ii)] $\delta_{\eta}$ is a derivation operator: 
$\delta_\eta(fg)=(\delta_\eta f)g+f(\delta_\eta g)$.
\item[(iii)] $\delta_\eta$ inserts a variable into the correlation functions:
\beq \label{eq:insertion-relation} \delta_\eta W_n(x_1,\dots,x_n)=W_{n+1}(x_1,\dots,x_n,\eta).\eeq
This property is equivalent to impose that 
\beq
\delta_\eta K(x_1,x_2)=-K(x_1,\eta)K(\eta,x_2).
\eeq
\item[(iv)] $\delta_{\eta}M(x)$ is of order $\hbar$, and is expressed in terms of $M(x)$, $M(\eta)$ and their $t$-derivatives. 
\item[(v)] $\delta_{\eta}$ commutes with $\partial_{t}$.
\end{itemize}
With these properties it is then possible to show that the $\hbar$ expansion of $W_n$ must start at least at $\hbar^{n-2}$: Firstly, Property \eqref{eq:insertion-relation} implies 
\beq W_{n}(x_{1},\dots,x_{n}) = \delta_{x_{n}}\cdots\delta_{x_{3}}W_{2}(x_{1},x_{2}).\eeq 
On the other hand, Properties (iv) and (v) imply 
\beq \label{eq:deldelM-is-On} 
\delta_{\eta_{1}}\cdots\delta_{\eta_{n}}M(x) = O(\hbar^{n}). \eeq
Then, since $W_{2}(x_{1},x_{2}) = O(\hbar^{0})$ is expressed by $M$ (see \eqref{alternative}), we get the desired property \eqref{eq:order-n-2-Wn}.

In \cite{BBEnew} and \cite{P5} explicit formulas are proposed for the definition of a suitable insertion operator through its action on the solution of the isomonodromy system:
\beq \label{eq:insertion-operator-def} \delta_{\eta} \Psi(x)= \left(\frac{M(\eta)}{x-\eta}+Q(\eta)\right)\Psi(x), \eeq
where $Q(\eta)$ is a matrix that depends on the Lax pair and is determined by imposing the above properties. Note that, the insertion operator \eqref{eq:insertion-operator-def} satisfies \eqref{eq:insertion-relation} for any choice of $Q(\eta)$. 
Property (iv) requires 
\beq \label{eq:order-1-for-M} \delta_{\eta} M(x) = - \frac{[M(x), M(\eta)]}{x-\eta} + [Q(\eta), M(x)] = O(\hbar). \eeq
Moreover, the condition (v) implies 
$[\delta_{\eta}, \partial_{t}] \Psi(x) = 0$; namely,
\begin{equation} \label{eq:commutativity-at-level-1}
\delta_{\eta} {\mathcal R}(x) = \hbar \partial_{t} Q(\eta) + [Q(\eta), {\mathcal R}(x)] + \left[ M(\eta), \frac{{\mathcal R}(x) - {\mathcal R}(\eta)}{x-\eta} \right].
\end{equation}
Condition \eqref{eq:commutativity-at-level-1} almost determines the action of $\delta_{\eta}$, as we explain below. 

Now let us consider the JM Lax pair to explain how the insertion operator is specified. First, look at the equality \eqref{eq:commutativity-at-level-1}. The l.h.s. given by 
\[
\delta_{\eta} {\mathcal R}(x) =
\begin{pmatrix} 
\frac{1}{2}\delta_\eta q & 0 \\ 
- \delta_\eta p & - \frac{1}{2} \delta_\eta q
\end{pmatrix}
\]
is independent of $x$. (The action of $\delta_\eta$ on $q$ and $p$ will be defined later so that \eqref{eq:commutativity-at-level-1} is satisfied; see \eqref{eq:consequence-for-delta-q} and \eqref{eq:consequence-for-delta-p} below.) Since ${\mathcal R}(x)$ is linear in $x$, the only $x$-depending term in the r.h.s. of \eqref{eq:commutativity-at-level-1} is 
\begin{multline*}
[Q(\eta), {\mathcal R}(x)] = \\
{\small 
\begin{pmatrix}
- p\,Q(\eta)_{1,2}-\frac{1}{2}Q(\eta)_{2,1} &
\frac{1}{2}\bigl( Q(\eta)_{1,1} -Q(\eta)_{2,2} \bigr) -(x+q)Q(\eta)_{1,2} \\
p\,\bigl( Q(\eta)_{1,1} - Q(\eta)_{2,2} \bigr)  -(x+q)Q(\eta)_{2,1} & 
p\,Q(\eta)_{1,2}+\frac{1}{2}Q(\eta)_{2,1} 
\end{pmatrix}
}
\end{multline*}
Therefore, the $x$-independence requires 
\begin{equation} \label{eq:requirement-for-Q-1}
Q(\eta)_{1,2} = Q(\eta)_{2,1} = 0,
\end{equation} 
and \eqref{eq:commutativity-at-level-1} is reduced to the following set of equalities:
\begin{eqnarray} 
\label{eq:requirement-for-Q-2}
2M(\eta)_{1,2} & = & Q(\eta)_{1,1}-Q(\eta)_{1,2}, \\[+.1em] 
\label{eq:consequence-for-delta-q}
\delta_{\eta} q & = & 2\hbar\,Q(\eta)_{1,1}, \\[+.1em] 
\label{eq:consequence-for-delta-p}
\delta_\eta p & = & - 2 \hbar \partial_t M(\eta)_{1,1}. 
\end{eqnarray}
Here we have used the following equalities 
(Cf. \eqref{eq:differential-system-for-projector}): 
\bea  \label{eq:differential-system-for-projector-component-wise}
\hbar \partial_t M(x)_{1,1}
& = & \frac{1}{2}M(x)_{2,1}+p M(x)_{1,2},\cr
\hbar\partial_t M(x)_{1,2}
& = &
(x+q)M(x)_{1,2}-M(x)_{1,1}+\frac{1}{2},\cr
\hbar\partial_t M(x)_{2,1}
& = &
-2p M(x)_{1,1}-(x+q)M(x)_{2,1}+p,\cr
\hbar\partial_t M(x)_{2,2}&=&
-\hbar\partial_t M(x)_{1,1}. 
\eea
Thanks to \eqref{eq:differential-system-for-projector-component-wise}, if we take $Q(\eta)$ satisfying \eqref{eq:requirement-for-Q-1} and \eqref{eq:requirement-for-Q-2}, straightforward computations shows that:
\bea \label{deltaM2}
\delta_\eta M(x)_{1,1}&=&-\frac{2\hbar}{x-\eta}\left( M(x)_{1,2}\partial_tM(\eta)_{1,1}-M(\eta)_{1,2}\partial_tM(x)_{1,1}\right),\cr
\delta_\eta M(x)_{1,2}&=&-\frac{2\hbar}{x-\eta}\left( M(x)_{1,2}\partial_tM(\eta)_{1,2}-M(\eta)_{1,2}\partial_tM(x)_{1,2}\right),\cr
\delta_\eta M(x)_{2,1}&=&\frac{\hbar}{p(x-\eta)}\left( M(x)_{2,1}\partial_tM(\eta)_{2,1}-M(\eta)_{2,1}\partial_tM(x)_{2,1}\right)\cr
&&-\frac{2\hbar}{p} M(x)_{2,1}\partial_tM(\eta)_{1,1},\cr
\delta_\eta M(x)_{2,2}&=&-\delta_\eta M(x)_{1,1}. 
\eea
which are obviously of order $O(\hbar)$. As a conclusion, in our JM case we can take:
\beq Q(\eta)=M(\eta)_{1,2}
\begin{pmatrix}1&0\\0&-1\end{pmatrix}, 
\eeq
\beq \label{eq:del-q-p} \delta_{\eta} q = 2 \hbar \partial_{t}M(\eta)_{1,2}, \quad\delta_{\eta} p = -2 \hbar \partial_{t}M(\eta)_{1,1}. \eeq
Then, the important conditions \eqref{eq:order-1-for-M} and \eqref{eq:commutativity-at-level-1} are satisfied.

From \eqref{deltaM2} it is tempting to conclude that the insertion operator $\delta_\eta$ satisfies \eqref{eq:deldelM-is-On}. However \eqref{deltaM2} is not sufficient to prove \eqref{eq:deldelM-is-On} due to the following reason. Since the r.h.s. of \eqref{deltaM2} involves time derivatives of $M(x)$, iterative application of the insertion operators creates terms of the form $\delta_\eta \partial_t M(x)$, $\delta_\eta \partial_{t}^{2}M(x)$ and so on. The problem is that the condition \eqref{eq:commutativity-at-level-1} is not enough to prove $[\delta_{\eta}, \partial_{t}] = 0$ as operators acting on PV ring. For example, the condition \eqref{eq:commutativity-at-level-1} is not enough to prove $[\delta_{\eta},\partial_{t}^{2}] \Psi(x) = 0$. To have this identity, we also need to require 
\beq \label{eq:requirement-for-insertion-operator-in-higher-t-derivation}
[\delta_{\eta},\partial_{t}]{\mathcal R}(x) = 0,
\eeq
and previous works never checked this identity. In the JM case, this condition is equivalent to $[\delta_\eta, \partial_t]q = [\delta_\eta, \partial_t]p = 0$. Since $\hbar \partial_t q = p + q^2 + \frac{t}{2}$, straightforward computation shows
\begin{eqnarray*}
\hbar [\delta_\eta, \partial_t]q 
& = & 
\delta_\eta p + 2q\, \delta_\eta q + \frac{1}{2} \delta_\eta t 
- \hbar \partial_t \delta_\eta q \\
& = & 
- 2 \hbar \partial_t M(\eta)_{1,1} + 4q\, \hbar \partial_t M(\eta)_{1,2}
+ \frac{1}{2} \delta_\eta t - 2 \hbar^2 \partial_t^2 M(\eta)_{1,2}.
\end{eqnarray*}
Thus we need to further require 
\begin{equation} \label{eq:delta-eta-t}
\delta_\eta t = 4 \hbar \partial_t M(\eta)_{1,1} 
- 8q\,\hbar \partial_t M(\eta)_{1,2} + 4 \hbar^2 \partial_t^2 M(\eta)_{1,2}
\end{equation}
to have $[\delta_\eta, \partial_t]q = 0$. The equality \eqref{eq:delta-eta-t} contradicts to Condition (v) since the r.h.s. of \eqref{eq:delta-eta-t} depends on $t$. Indeed, it follows from \eqref{M0JM} that the first term of its $\hbar$-expansion is given by 
\begin{eqnarray*}
\delta_\eta t & = & \hbar \Bigl( 4 \partial_t M^{(0)}(\eta)_{1,2} 
- 8 q_0 \partial_t M^{(0)}(\eta)_{1,1} \Bigr) + O(\hbar^2) \\
& = & 
- \hbar \frac{\eta + q_0}
{\bigl( (\eta+q_0)^2 + \frac{\theta}{q_0} \bigr)^{3/2}}
+O(\hbar^2).
\end{eqnarray*}
This implies that $[\delta_\eta, \partial_t]t \ne 0$ which proves that the insertion operator $\delta_\eta$ does not satisfy Condition (v). Consequently this method does not prove \eqref{eq:deldelM-is-On} and the leading order property of $W_n$. 

Unfortunately this problem is not specific to the JM Lax pair and is really intrinsic to the current method of the insertion operator. For example it also arises in the HTW Lax pair as well. We could not find any simple way to fix the problem and it is likely that substantial modifications of the insertion operator are required. However since the insertion operator exists in the context of random matrix models, we believe that it should exist in the context of determinantal formulas too. 


\section{\label{AppendixD}Tau-function and symplectic invariants}

Here we describe a relation between the tau-function \eqref{eq:isomonodromic-tau-function} and the (generating function of) symplectic invariants, following the idea of \cite{BBEnew, P2WithoutCoeffFull}.

\begin{thm}[{Theorem 5.1 of \cite{EO}}] 
For $g \ge 1$, both $F^{(g)} = F_{\rm JM}^{(g)}$ and $F_{\rm HTW}^{(g)}$ satisfy 
\beq \label{eq:t-derivation-Fg}
\frac{dF^{(g)}}{dt} = 2 \Res_{x = \infty} \left( \frac{\partial s_{\infty}}{\partial t}(x) W^{(g)}_{1}(x)dx \right).
\eeq
Here $W_{1}(x)$ and $s_{\infty}(x)$ are given in \eqref{eq:W1} and \eqref{eq:s-infty-JMHTW}, respectively.
\end{thm}

\begin{proof}
The fact that the JM and HTW Lax pair are of topological type and
the result of Proposition \ref{Conditions} (i) imply  
\begin{equation}
W^{(g)}_{1}(x(z))dx(z) = \omega^{(g)}_{1}(z),
\end{equation} 
where $x(z)$ appears in the parametrization \eqref{SpecCurveJM2} or \eqref{SpecCurveHTW2} of the spectral curve, and $\omega^{(g)}_{1}(z)$ is the Eynard-Orantin differential of type $(g,1)$. On the other hand, the function 
\[
\Lambda(z) = \frac{\partial s_{\infty}}{\partial t}(x(z))
\] 
satisfies \smallskip
\[
\hspace{-20.em}
\frac{\partial x}{\partial t}(z) dy(z) - 
\frac{\partial y}{\partial t}(z) dx(z)
\]
\vs{-1.2em}
\bea 
= \left\{
\begin{array}{ll}
\displaystyle \Res_{w=\infty} \Lambda(w)\omega^{(0)}_{2}(w,z) - \Res_{w=0} \Lambda(w)\omega^{(0)}_{2}(w,z) & \text{for JM case}, 
\\[+.7em]
\displaystyle \Res_{w=\infty} \Lambda(w)\omega^{(0)}_{2}(w,z) & \text{for HTW case}.
\end{array}
\right.
\eea
This is the required condition for $\Lambda(z)$ to apply Theorem 5.1 of \cite{EO} which proves \eqref{eq:t-derivation-Fg}. This completes the proof of Theorem \ref{propo1} and Theorem \ref{propo2}. 
\end{proof}

\section{Constant term in \eqref{eq:difference-Fg-intro} and the symplectic invariance} 
\label{AppendixE} 
The symplectic transformation \eqref{eq:Yamada-symplectic-transform} from JM curve to HTW curve is obtained from the composition of following three symplectic transformations: 
\begin{eqnarray}
\label{eq:symp-1}
(x_1, y_1) = (x_{\rm JM}, y_{\rm JM})  \mapsto  (x_{2}, y_{2}) &  \text{by}  &  x_2 = x_1,~~y_2 = y_1 + x_1^2 + \frac{t}{2} \hspace{+3.em} \\
\label{eq:symp-2}
(x_2, y_2) \mapsto (x_3, y_3) &  \text{by}  & x_3 = y_2,~~y_3 = - x_2 \\
\label{eq:symp-3}
(x_3, y_3) \mapsto (x_4, y_4) =  (x_{\rm HTW}, y_{\rm HTW})  &  \text{by}  & x_4 = x_3,~~y_4 = y_3 + \frac{\theta}{2x_3}.
\end{eqnarray}
Since the symplectic transformations \eqref{eq:symp-1} and \eqref{eq:symp-3} preserve the branch points and the recursion kernel $K(z_0, z)$, we can conclude that the free energy $F^{(g)}$ is invariant (Cf. Theorem 7.1 in \cite{EO}). On the other hand, the second symplectic transformation \eqref{eq:symp-2} does not preserve $F^{(g)}$ in general as shown in \cite{Bouchard-Sulkowski, EO-xy-symmetry}, and Theorem 3.1 in \cite{EO-xy-symmetry} proves that an integral of $\omega^{(g)}_1$ appears as their difference. In our case Theorem 3.1 in \cite{EO-xy-symmetry} shows
\beq 
F_{\text{JM}}^{(g)}-\frac{\theta}{2-2g}\int_0^{\infty}\omega_{\text{JM},1}^{(g)}=F_{\text{HTW}}^{(g)}+\frac{\theta}{2(2-2g)}\int_{-1}^{1}\omega_{\text{HTW},1}^{(g)} 
\quad \text{for $g \ge 2$}. \eeq
This is equivalent to say that:
\beq \label{eq:correct-symplectic-invariance-JMHTW}
F_{\text{JM}}^{(g)}-F_{\text{HTW}}^{(g)}=\frac{\theta}{2-2g}\left(\int_0^{\infty}\omega_{\text{JM},1}^{(g)}+\frac{1}{2}\int_{-1}^{1}\omega_{\text{HTW},1}^{(g)}\right) \quad \text{for $g \ge 2$}. \eeq
This relation for $g\in \{2,3\}$ can be directly tested from our computations. In fact from our work in Section \ref{section:Limit-JM-to-Weber} and Section \ref{section:Limit-HTW-to-Bessel}, we obtained the exact value of the r.h.s. of \eqref{eq:correct-symplectic-invariance-JMHTW} (Cf. \eqref{eq:JM-HTW-Section-4}). Moreover, it is straightforward to observe that $\omega_{\text{JM},1}^{(g)}\underset{q_0\to 0}{\to}0$ and $\omega_{\text{HTW},1}^{(g)}\underset{q_0\to +\infty}{\to}0$ for all $g\geq 2$. Therefore, we obtain:
\beq 
\int_0^\infty \omega_{\text{Weber},1}^{(g)}=\frac{B_{2g}}{2g\,\theta^{2g-1}} \text{ ~ and ~ } \int_{-1}^1 \omega_{\text{Bessel},1}^{(g)}=\frac{B_{2g}}{g\,\theta^{2g-1}}\quad \text{for $g\geq 2$.}\eeq
In both cases, the choice of integration contour does not matter since the differential forms do not have any residue. The only requirement is that the contours avoid $\{\pm 1\}$ in the Hermite-Weber case and $\{0\}$ in the Bessel case.



\section*{Acknowledgment}
The authors would like to thank 
Ga\"etan Borot, 
Bertrand Eynard, 
Alba Grassi,
Masahide Manabe, 
Motohico Mulase, 
Yasuhiko Yamada 
for fruitful discussion.
O.~Marchal would like to thank Universit\'e Lyon $1$ 
and particularly Universit\'e Jean Monnet and 
Institut Camille Jordan for the opportunity to make 
this research possible. O.~Marchal would also like 
to thank his family and friends for moral support 
during the preparation of this article.
K.~Iwaki is supported by 
JSPS KAKENHI Grant Number 13J02831, 16K17613 and 16H06337. 
O.~Marchal is supported by the LABEX MILYON (ANR-10-LABX-0070) 
of Universit\'e de Lyon, within the program 
``Investissements d'Avenir'' (ANR-11-IDEX-0007) 
operated by the French National Research Agency (ANR). 
The authors also grateful to Springer International Publishing.
The final publication is available at Springer via 
\href{http://dx.doi.org/10.1007/s00023-017-0576-z}
{http://dx.doi.org/10.1007/s00023-017-0576-z}.

\end{document}